\documentclass[
superscriptaddress,
reprint,
 amsmath,amssymb,
 aps,
 twocolumn
pra
]{revtex4-2}

\usepackage{graphicx}
\usepackage{dcolumn}
\usepackage{bm}

\usepackage{mathtools}
\usepackage{physics}
\usepackage{amsmath}
\usepackage{bm}
\usepackage{CJKutf8}
\usepackage{subcaption}
\usepackage[justification=raggedright, singlelinecheck=false]{caption}
\usepackage[normalem]{ulem}  
\usepackage{xcolor}
\usepackage{tikz}
\usetikzlibrary{calc}

\usepackage{xcolor}

\usepackage{booktabs}
\usepackage{multirow}
\usepackage{array}
\usepackage{hhline}
\usepackage{braket}
\usepackage[bookmarks=false,
 breaklinks=true,pdfborder={0 0 0},pdfborderstyle={},backref=false,colorlinks=true]
 {hyperref}

\begin{document}
\title{Capacity-time Trade-off in Highly Reliable Quantum Memory}
\author{Miao-Miao Yi (\begin{CJK}{UTF8}{gbsn}易淼淼\end{CJK})}
\affiliation{Graduate School of China Academy of Engineering Physics, Beijing, 100193, China}
\author{L. X. Cui (\begin{CJK}{UTF8}{gbsn}崔廉相\end{CJK})}
\affiliation{Beijing Computational Science Research Center, Beijing, 100193, China}
\affiliation{International Research Center for Neurointelligence, University of Tokyo Institutes for Advanced Study, University of Tokyo, Bunkyo-ku, Tokyo, 113-0033, Japan}
\author{Y -M. Du (\begin{CJK}{UTF8}{gbsn}杜亦牧\end{CJK})}
\email{ymdu@gscaep.ac.cn}
\affiliation{Graduate School of China Academy of Engineering Physics, Beijing, 100193, China}
\author{C. P. Sun (\begin{CJK}{UTF8}{gbsn}孙昌璞\end{CJK})}
\email{suncp@gscaep.ac.cn}
\affiliation{Graduate School of China Academy of Engineering Physics, Beijing, 100193, China}
\begin{abstract}
Reliable quantum storage in practice relies on precise calibration of key parameters, notably the global detuning, while inevitably being subject to the combined influence of multiple disorder sources. In this work, a comprehensive model for an Electromagnetically induced transparency (EIT) protocol is considered, in which coupling disorder and detuning disorder are incorporated simultaneously. After quantitatively analyzing the control dynamics, a highly precise phase–detuning relation to improve calibration accuracy is obtained. Building on this result, a Berry-phase-based control strategy is proposed to mitigate the degradation caused by the global detuning. We further reveal that there exists a joint effect simultaneously induced by different disorder sources, which can substantially reshape the decoherence. Finally, an effective notion of storage capacity is introduced and a general time–capacity relation is obtained, providing guidance for subsequent experimental optimization and device design.
\end{abstract}
\maketitle
\section{Introduction}
Quantum memory is a system that can store and retrieve a quantum state on demand, which holds significant importance for advancing numerous fields including quantum communication \cite{gisin2007quantum_communication,scarani2009security_QKD,
lvovsky2009optical_quantum_memory} and quantum computation \cite{mermin2007quantum_computation_science,shor1995quantum_computing,lvovsky2009optical_quantum_memory}. A high-quality quantum storage must satisfy several essential criteria:  (1) a long storage time \cite{lvovsky2009optical_quantum_memory,RZhao2009long_lived_quantum_memory,LPYang2023storage_efficiency}, (2) large capacity and universality for storing arbitrary states \cite{hashimoto2019,CPSun2003quasi,lvovsky2009optical_quantum_memory}, and (3) a perfect store-in-retrieve-out (SIRO) process without destructing information. Driven by these requirements, many theoretical schemes have been proposed, particularly those based on atomic ensembles \cite{lukin2000quantum_memory,fleischhauer2002quantum_memoryPRA,CPSun2003decoherence,Tomography2008squeezed}.

In realistic implementations, the reliable operation of quantum memories is affected by several nonideal factors, which calls for accurate performance characterization and optimization. On the one hand, unavoidable global detuning, arising for example from AC-Stark shifts \cite{Tomography2008squeezed, HYF2018ac_stark_shift_storage_efficiency,ACstarkshift2009,ACstarkshift2023}, typically introduces an additional overall phase shift to the stored state, so precise operation of a memory depends critically on accurate calibration of the relevant experimental parameters \cite{Tomography2008squeezed,Tomography2009,Tomography2013}. On the other hand, multiple disorder sources are inevitable, including inhomogeneous light-atom coupling\cite{CPSun2003decoherence,Kimiaee2025DisorderedCoupling,Gallego2018_spatial_light}, linewidth-induced detuning disorder\cite{Cheng2020LaserLinwidth}, and inhomogeneous broadening \cite{Greentree2009Broadening,Eberly_doppler_broadening1998}. These disorder effects mainly introduce random phase noise into the stored state. In addition, other mechanisms, such as atomic motion \cite{fleischhauer2002quantum_memoryPRA}, spontaneous emission \cite{LPYang2023storage_efficiency,Spontaneous2007PRA,Spontaneous2013PRL}, four-wave mixing \cite{Four-wave-mixing2013,Four-wave-mixing2014,Four-wave-mixing2019}, and spin-wave decay \cite{Spin-wave2015PRL,Spin-wave2019PRL}, can cause irreversible loss and decoherence, and can further affect the storage lifetime in many specific implementations. Accordingly, a number of studies have addressed the characterization and mitigation of such effects \cite{LPYang2023storage_efficiency,Spontaneous2013PRL,Four-wave-mixing2019,Four-wave-mixing2014,Spin-wave2015PRL,lvovsky2009optical_quantum_memory}.

Nevertheless, several key issues still merit further study. First, calibration accuracy remains to be improved. In experiments \cite{Tomography2008squeezed,Tomography2009,Tomography2013}, the phase induced by an unknown global detuning is usually measured by tomography and then converted into a detuning value through an empirical phase-detuning relation. Any systematic error in this inference could be amplified over long storage times. Second, when different nonideal factors coexist, they could produce complex joint effects on the memory, that cannot be fully captured by conventional analyses treating each imperfection in isolation, and this in turn can prevent a full account of their impact on storage performance. These issues motivate a unified treatment of multiple imperfections, a more accurate phase-detuning relation for calibration, and effective strategies for improving storage performance.

It is also worth emphasizing that ensemble-based quantum memories, including EIT protocols, are in principle capable of storing arbitrary optical states supported within the Fock subspace allowed by the available collective atomic excitations \cite{lukin2000quantum_memory,fleischhauer2002quantum_memoryPRA,CPSun2003quasi,hashimoto2019}. For ensembles with a large atom number, this supported subspace can be large, and such memories could thus serve as feasible universal quantum memories. Meanwhile, there is growing interest in using nonclassical optical states distributed over many Fock components, such as cat states, as resources for more general quantum networks and quantum computation \cite{GKP2024,CAT2013,coherent_state2003PRA,Qudit2018,hashimoto2019}. This motivates going beyond the usual single-photon setting \cite{LPYang2023storage_efficiency,Philipp2022singlePhotonState,hashimoto2019} and considering the degradation of more general multiphoton storage. Since such states typically occupy a larger Fock subspace and therefore carry more information, it is necessary to introduce an effective measure of information-storage capacity for a universal memory. Qualitatively, a larger support subspace of the stored state may lead to more pronounced disorder-induced dephasing from disordered coupling and disordered detuning, and related imperfections. This in turn motivates a quantitative study of these effects on the storage capacity, as well as its trade-off with storage time, to characterize storage performance and provide guidance for experiments.

In this paper, within the framework of quantum reliability \cite{reliability2023Cui,reliability2025Cui,reliability2025Du,reliability2025Du2}, we consider a comprehensive model for the EIT-based protocol \cite{CPSun2003quasi,lukin2000quantum_memory,fleischhauer2002quantum_memoryPRA} to simultaneously investigate the effects of disordered coupling strength and detuning on quantum memory and the key operational processes in memory-based devices. Here, these disorders are modeled mainly as static inhomogeneous broadening \cite{Greentree2009Broadening,Eberly_doppler_broadening1998} and static disordered coupling \cite{CPSun2003decoherence}. We quantitatively characterize the performance degradation induced by random Berry's phase arising from these practical imperfections. This analysis yields a more precise phase-detuning relation for calibrating the global detuning. Building on this calibrated relation, a Berry's phase-based control strategy is provided to effectively mitigate degradation of the memory. Furthermore, we show that the impact of disordered coupling can be dominated by its correlation with detuning, rather than individually, such correlation effects impose constraints on the driving time of the control pulse. This generalizes the conventional degradation analysis beyond in-isolation treatments. Moreover, we introduce an effective notion of storage capacity, and obtain a general time-capacity trade-off that can inform experimental parameter optimization. 

This paper is organized as follows. In Sec.~\ref{Section2_letter}, we introduce the physical model of the memory and present the definition of storage capacity. In Sec.~\ref{Section3_letter}, we present the actual evolution of stored states due to imperfections, formulate the quantum reliability of the quantum memory and memory based devices, and we reveal the correlation effects among different disorders. In Sec.~\ref{Section4_letter}, we provide a more precise phase-detuning relation for calibration. In Sec.~\ref{Section5_letter}, we present the effects of disorders and reveal a universal constraint on stored states. In Sec.~\ref{Section6_letter}, we present the capacity-time trade-off. In Sec.~\ref{Section7_letter}, we summarize our results.

\section{Model}\label{Section2_letter}
\begin{figure}
    \centering
    \includegraphics[width=1\linewidth]{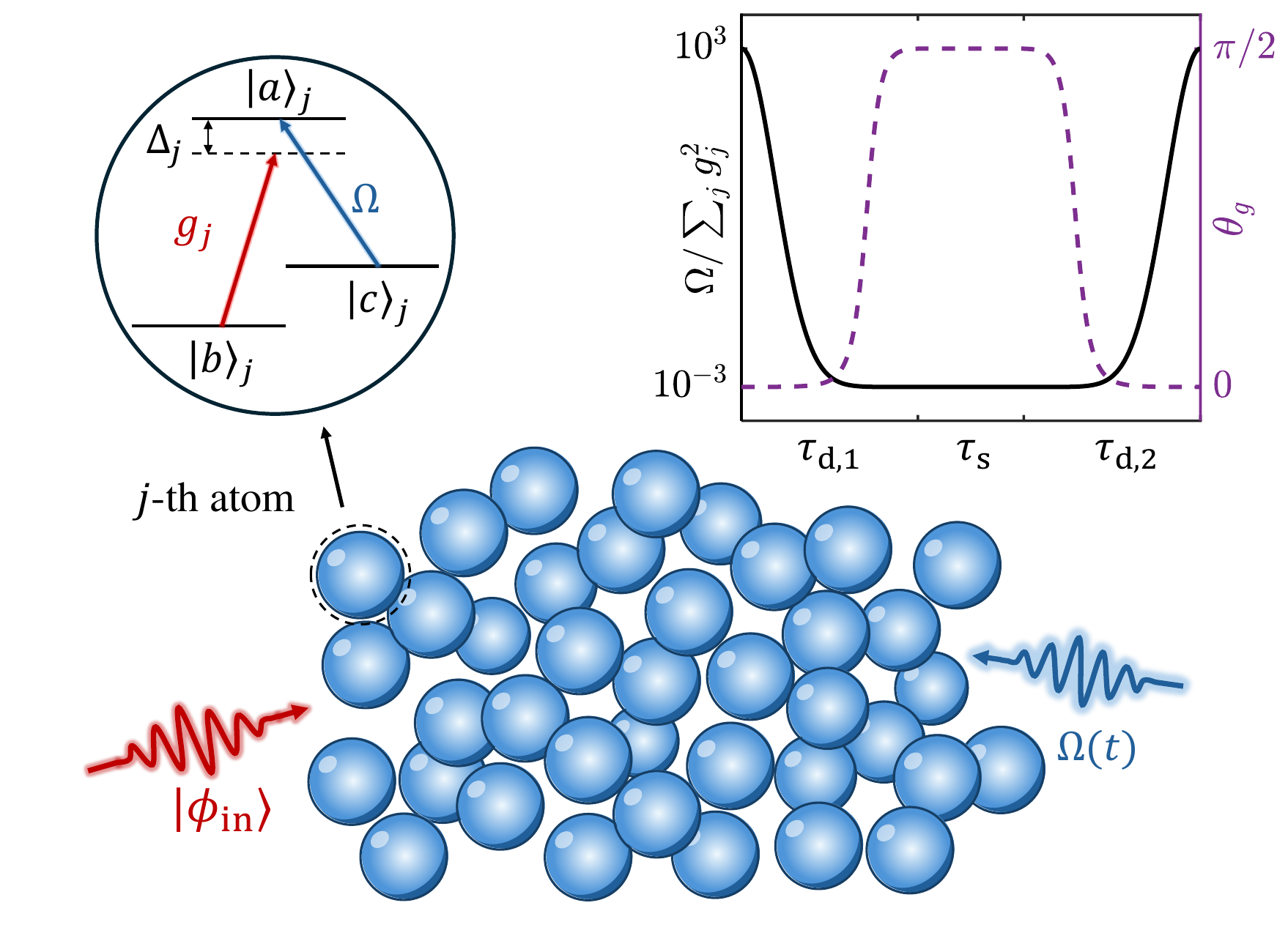}
    \caption{Schematic of a quantum memory system composed of $N$ three-level atoms, where atoms are coupled to a classical control field (blue) and a quantum mode (red). The upper-right panel illustrates the classical driving pulse $\Omega(t)$ (black solid line) and the corresponding angle $\theta_{\bm g}(t)$ (purple dashed line) over three time intervals to realize a SIRO process. Here, a Gaussian pulse is taken as an example, $\Omega/\sum_jg_j^2=\xi\exp(-2\ln(\xi)(t/\tau_{\mathrm{d},1})^2)$ for store-in, with the inverse procedure used for retrieval, here we set $\xi=1000$.}
    \label{fig1_letter}
\end{figure}

The system we consider consists of $N$ $\Lambda$-type three-level atoms shown in Fig.~\ref{fig1_letter}. For the $j$-th atom, the ground state, excited state and metastable state are labeled by $|b\rangle_j$, $|a\rangle_j$ and $|c\rangle_j$, respectively. The $|b\rangle_j \rightarrow|a\rangle_j$ transition couples to a quantized mode that carries information, with coupling strength $g_j$, while $|c\rangle_j \rightarrow|a\rangle_j$ couples to a classical control field with Rabi frequency $\Omega_j$. In the interaction picture, the Hamiltonian follows as ($\hbar=1$)
\begin{align}
    \hat H=&\,\hat a\sum_{j=1}^N g_j \exp(\mathrm{i}\Delta_j t)\exp(\mathrm{i}\bm K_{ba}\cdot\bm r_j)\hat\sigma_{ab}^{(j)}\label{Hamiltonian_initial_letter}\\&\,+\sum_j^N\Omega_j(t)\exp(\mathrm{i}\delta_jt)\exp(\mathrm{i}\bm K_{ca}\cdot \bm r_j)\hat\sigma^{(j)}_{ac}+\mathrm{H.c.},\notag
\end{align}
where $\bm r_j$ represents the position of the $j$-th atom, $\bm K_{ba} (\bm K_{ca})$ is the wave vector of the quantum (classical) field, $\hat a$ is the annihilation operator of the quantized mode, and $\hat\sigma^{(j)}_{\mu\nu}=|\mu\rangle_{jj}\langle\nu|$ ($\mu,\nu=a,b,c$) is the quasi-spin operator described for the transition $|\nu\rangle_j \rightarrow|\mu\rangle_j$. Here, $\Delta_j\ (\delta_j)$ represents the ‌unavoidable detuning between the quantum (classical) field and the $j$-th atom. To simplify the discussion while capturing the essential physics, we consider $\delta_j\equiv0$, and $\Omega_j(t)\equiv\Omega(t)$, $\forall j$. Furthermore, we model the detuning and coupling strength of each atom as independent normally distributed random variables,  $\Delta_j\sim\mathcal{N}(\Delta,\delta\Delta^2)$ and  $g_j\sim\mathcal{N}(g,\delta g^2)$, $\forall j$, thereby modeling static inhomogeneous broadening and static disordered coupling in the ensemble \cite{Eberly_doppler_broadening1998,Greentree2009Broadening,CPSun2003decoherence}. Here, $\Delta$ and $\delta \Delta$ represent the mean and broadening of the local detuning, which could arise from position-dependent AC-stark \cite{Tomography2008squeezed,ACstarkshift2023} and Doppler shifts \cite{Eberly_doppler_broadening1998}, while $g$ and $\delta g$ denote the mean and disordered coupling strength, with the latter reflecting local disorder induced by impurities \cite{CPSunAndZSong2005spin,CPSun2003decoherence} or spatial inhomogeneity of the optical field \cite{Gallego2018_spatial_light}.

We consider $N\gg1$ and the low excitation condition \cite{CPSun2003quasi,CPSunAndZSong2005spin,CPSun2004Noabelian_geometry_quantum_memory}, where the occupation number in the excited (metastable) state within the atomic ensemble is negligible compared to $N$. In this case, all eigenstates of Hamiltonian in Eq.~(\ref{Hamiltonian_initial_letter}) can be obtained (please see Appendix \ref{Section 1_SM} for details), and we focus on the zero-eigenvalue dark states employed in quantum memory \cite{CPSun2003quasi,lukin2000quantum_memory,fleischhauer2002quantum_memoryPRA}
\begin{equation}
    |D_{\bm \xi,n}\rangle=\frac{1}{\sqrt{n!}}(\hat D_{\bm \xi}^\dagger)^n|\bm 0\rangle,\label{dark-state-letter}
\end{equation}
where the ground state of the system in Eq.~(\ref{Hamiltonian_initial_letter}) is defined by $|\bm 0\rangle=|0\rangle_L\otimes|0\rangle_A$ with $\hat H|\bm 0\rangle=0$, where $|0\rangle_L$ is the vacuum of quantum field, and $|0\rangle_A:=\prod_j|b\rangle_j$ denotes the ground state of the atomic ensemble. The dark-state excited operator is defined as
\begin{equation}
    \hat D_{\bm \xi}:=\hat a\cos\theta_{\bm g}-\hat C_{\bm \xi}\sin\theta_{\bm g},
\end{equation}
with relationships $[\hat D_{\bm\xi},\hat D^{\dagger}_{\bm\xi}]=1$ and $[\hat H,\hat D_{\bm\xi}]=0$, where $\tan\theta_{\bm g}:=(\sum_j g_j^2)^{1/2}/\Omega(t)$. Here, the quasi-spin-wave collective excited operator is written as
\begin{equation}
    \hat C_{\bm \xi} =\sum_j^N\frac{g_j}{(\sum_k g_k^2)^{1/2}}\hat\sigma_{bc}^{(j)}\exp(-\mathrm{i}\bm K_{bc}\cdot\bm r_j-\mathrm{i}\Delta_jt),
\end{equation}
where subscript $\bm\xi=\{\bm\Delta,\bm g\}$ denotes a set of inhomogeneous detuning $\bm\Delta=(\Delta_1,\Delta_2,\cdots,\Delta_N)$ and coupling $\bm g=(g_1,g_2,\cdots,g_N)$, and the wave vector $\bm K_{bc}=\bm K_{ba}-\bm K_{ca}$ is introduced to depict the second order transition $|b\rangle\rightarrow|c\rangle$. Here, further details can be found in Appendix \ref{Section 1_SM}.

It is observed from Eq.~(\ref{dark-state-letter}) that $|D_{\bm\xi,n}\rangle=|n\rangle_L\otimes|0\rangle_A$ when $\theta_{\bm g}=0$, and $|D_{\bm\xi,n}\rangle=(-1)^n|0\rangle_L\otimes|n\rangle_A$ when $\theta_{\bm g}=\pi/2$. Therefore, as shown in Fig.~\ref{fig1_letter}, for a general optical stored state $|\phi_{\mathrm{in}}\rangle=\sum_n C_n |n\rangle_L$, the ideal SIRO process in the EIT-based protocol can be divided into three stages as follows. Before storage (i.e., $t=0$), we require the atomic ensemble to remain in $|0\rangle_A$, and tune the driving field to satisfy $\Omega(0)\gg(\sum_j g_j^2)^{1/2}$, so that $\theta_{\bm g}(0)=0$. During the driving interval $\tau_{\mathrm{d},1}$, the driving field is adiabatically tuned from $\Omega\gg(\sum_j g_j^2)^{1/2}$ to $\Omega\ll(\sum_j g_j^2)^{1/2}$, causing the angle $\theta_{\bm g}$ to vary adiabatically from $0$ to $\pi/2$. As a result, the information $\{C_n\}$ of the optical state can be mapped into the atomic ensemble as $|0\rangle_L\otimes\sum_n C_n (-1)^n |n\rangle_A$. The state then can be allowed to undergo a storage time, denoted as $\tau_{\mathrm{s}}$. Following this, 
during $\tau_{\mathrm{d},2}$, the state can be retrieved through the reversed process, in which $\theta_{\bm g}$ varies adiabatically from $\pi/2$ back to $0$. Notably, in this paper, we consider the symmetric SIRO protocol $\tau_{\mathrm{d},1}=\tau_{\mathrm{d},2}:=\tau_{\mathrm{d}}/2$, and focus on the relatively long storage time with $\tau_{\mathrm{s}}\gg\tau_{\mathrm{d}}$ for simplicity.

Moreover, to characterize the effective information-storage capacity of a universal quantum memory while retaining practical relevance, we introduce the maximized single-letter coherent information over the supported space \cite{holevo2019QI,nielsen2010QI},
\begin{equation}
    \mathcal{C}:=\max_{\rho\in\mathcal{H}_d}[S(\Lambda(\rho))-S(\mathcal{I}\otimes \Lambda(|\psi_{\rho}\rangle\langle \psi_{\rho}|))],
\end{equation}
and use it as an effective measure of the memory's information-storage capacity in this paper. Here, $S(\cdot)$ is the von Neumann entropy (e.g., $S(\rho)=-\mathrm{Tr}(\rho\ln\rho)$), $|\psi_\rho\rangle$ is a purification of $\rho$ and $\mathcal{H}_d$ denotes the fully supported $d$-dimensional Fock subspace, namely, the quantum memory is required to store arbitrary states in $\mathcal{H}_d$, thereby ensuring its universality. The channel $\Lambda(\cdot)$ denotes the quantum channel associated with a single SIRO process. Physically, the effective notion $\mathcal{C}$ characterizes the maximum amount of quantum information that can be reliably preserved by the memory over $\mathcal{H}_d$ in a single SIRO process. Further discussion of this effective information-storage capacity is given in Appendix \ref{Section 7_SM}.

\section{quantum reliability of quantum memory and memory-based devices}\label{Section3_letter}
In this section, we first derive the actual output state after a SIRO process in the simultaneous presence of detuning disorder and coupling disorder, and show that the stored state acquires an additional random Berry‘s phase. We then use this realistic evolution to formulate the quantum reliability of a quantum memory and of memory-based devices involving multiple SIRO processes, thereby quantifying how far their actual performance deviates from the ideal functionality. In particular, we show that reducing the correlations of disorders across different SIRO cycles can effectively suppress the reliability loss. Finally, we present one of the central findings of this paper: the impact of disordered coupling is not determined by disordered coupling alone, but rather is dominated by its joint effect with detuning disorder. This extends beyond the conclusion obtained in conventional analyses that treat individual imperfections in isolation.

\subsection{the actual evolution of the stored state}
We begin by presenting the actual output state of the memory in the presence of both disordered coupling and detuning. For a given inhomogeneous configuration $\bm\xi$, the output state is
\begin{equation}
    |\phi_{\mathrm{out},\bm{\xi}}\rangle=\sum_n C_n\exp(\mathrm{i}\gamma_{\bm{\xi},n}(\tau))|n\rangle_L,\label{final_state_letter}
\end{equation}
namely, the stored state acquires an additional Berry's phase throughout the entire SIRO process, with 
\begin{equation}
    \gamma_{\bm{\xi},n}(\tau)=-n\sum_j\frac{g^2_j\Delta_j}{\sum_k g_k^2}\int^\tau_0\mathrm{d}t\sin^2\theta_{\bm{g}}(t).\label{Berrys phase}
\end{equation}
The detailed derivations of the actual evolution are provided in Appendix \ref{SubSec.~2_SM}, and throughout this paper, the total evolution time is denoted as $\tau$, with $\tau\equiv\tau_{\mathrm{s}}+\tau_{\mathrm{d}}$. 

For a large atom number $N\gg1$, we here expand $\gamma_{\bm{\xi},n}(\tau)$ around the global (average) detuning $\Delta$ and coupling $g$ to first order
\begin{equation}
    \gamma_{\bm{\xi},n}(\tau)\simeq\gamma_{\bm{\xi}_0,n}(\tau)+\gamma_{\bm{\xi}_0,n}(\tau)\tilde{\varepsilon}_{\bm{\Delta}}+\mu_{\bm{\xi}_0,n}(\tau)\tilde\varepsilon_{\bm g},\label{Berry's_phase_expansion_letter}
\end{equation}
the subscript $\bm{\xi}_0:=\{\bm \Delta_0,\bm g_0\}$ denotes a set of homogeneous detuning $\bm{\Delta}_0=(\Delta,\Delta,\cdots,\Delta)$ and coupling strength $\bm{g}_0=(g,g,\cdots,g)$. Here,
\begin{equation}
    \mu_{\bm{\xi}_0,n}(\tau):=-\frac{n}{2}\Delta\int^\tau_0\mathrm{d}t\sin^2(2\theta_{\bm g_0}),\label{mu}
\end{equation}
and relative deviation 
\begin{equation}
    \tilde\varepsilon_{\bm{y}}:=(\sum_j y_j/N-y)/y,\quad y=\Delta,g.\label{epsilon_y_letter}
\end{equation}
Therefore, following from Eqs.~(\ref{Berry's_phase_expansion_letter}) and (\ref{epsilon_y_letter}), the output state in Eq.~(\ref{final_state_letter}) approximately depends solely on the collective detuning $\mathcal{D}:=\sum_j\Delta_j$ and the collective coupling $\mathcal{G}:=\sum_j g_j$.

\subsection{quantum reliability}\label{Quantum reliability}
As shown by the actual output state in Eq.~(\ref{final_state_letter}), the realistic SIRO process drives the system away from the ideal evolution. This could in turn degrade the functionality of the memory. Since quantum memory is directly employed in quantum devices such as synchronizer \cite{lvovsky2009optical_quantum_memory,synchronization2007} and repeaters \cite{repeaters2019Pirandola,repeaters2023Azuma,gisin2007quantum_communication}, which involve multiple storage operations, and different trajectories could mutually interfere. To assess the performance of quantum memory from a practical perspective, we introduce the framework of quantum reliability \cite{reliability2023Cui,reliability2025Cui,reliability2025Du,reliability2025Du2}, which quantifies the deviation between realistic and ideal processes. Here, the quantum reliability of two key operational procedures in such memory-based devices is mainly investigated.

\begin{figure}
    \centering
    \includegraphics[width=0.8\linewidth]{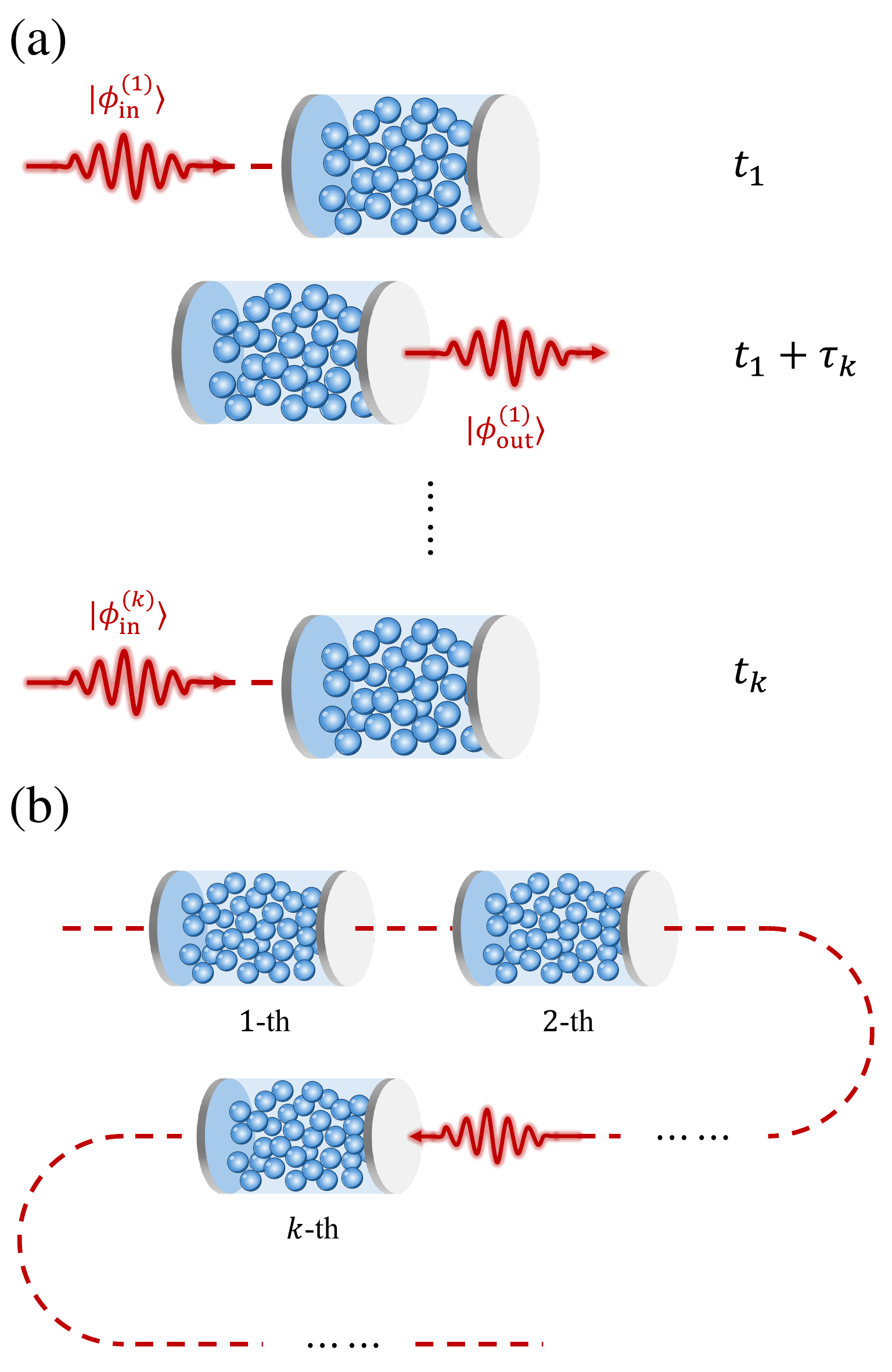}
    \caption{The key procedure of typical memory-based devices, i.e., the synchronizer and repeater chains: (a) the memory performs SIRO on a sequence of $k$ states, (b) a state undergoes $k$ memory-based repeaters.}
    \label{fig2_paper}
\end{figure}

\subsubsection{procedure (a) and its reliability}
As illustrated in Fig.~\ref{fig2_paper} (a), procedure (a) captures the key operation of a synchronizer in a quantum channel. The synchronizer is based on quantum memory that performs a SIRO process on a state in the channel to introduce a time delay, thereby synchronizing it with parallel states of other channels in subsequent operations \cite{synchronization2007,lvovsky2009optical_quantum_memory}. Here, the quantum memory is assumed to perform SIRO on a sequence of $k$ states $\{|\phi_{\mathrm{in}}^{(1)}\rangle,\cdots,|\phi_{\mathrm{in}}^{(k)}\rangle\}$, with $|\phi^{(j)}_{\mathrm{in}}\rangle=\sum_n C_n^{(j)}|n\rangle_L$. In the paper, we consider $\tau_j=\tau$, $\forall j$ to simplify the subsequent discussion. The quantum reliability of this procedure is written as a compact form
\begin{align}
    \mathcal{R}_a(k)
    =&\,\sum_{\bm n,\bm n'}|C(\bm n)|^2|C(\bm n')|^2\cos[\gamma_{\bm{\xi}_0,1}(\tau)\bm 1^{\mathrm{T}}(\bm n-\bm n')]\notag\\
    &\,\times\exp[-(\bm{n}-\bm{n'})^{\mathrm{T}}\bm 1\bm 1^{\mathrm{T}}(\bm{n}-\bm{n'})\Gamma_{\bm{\xi}_0}(\tau)/(2N)],\label{reliability_1_letter}
\end{align}
where
\begin{equation}
    \Gamma_{\bm{\xi}_0}(\tau)=(\delta\Delta/\Delta)^2\gamma_{\bm\xi_0,1}^2(\tau)+(\delta g/g)^2\mu^2_{\bm{\xi}_0,1}(\tau),\label{Gamma_letter}
\end{equation}
and $\bm{n}=(n_1,\cdots,n_k)^{\mathrm{T}}\in\mathbb{N}^k$, $\bm 1=(1,\cdots,1)^{\mathrm{T}}$, $C(\bm n)=\prod_j^kC^{(j)}_{n_j}$. For a detailed derivation of the quantum reliability of this procedure, please see Appendix \ref{subAppendix_Quantum_reliability_deviration}.

\subsubsection{procedure (b) and its reliability}
Fig.~\ref{fig2_paper} (b) shows the process of repeater chains in quantum network \cite{gisin2007quantum_communication,repeaters2019Pirandola,repeaters2023Azuma}, denoted as procedure (b), where an arbitrary optical state $|\phi_{\mathrm{in}}\rangle$ undergoes $k$ repeaters implemented by such memories. Here the collective coupling and detuning associated with different quantum memories are treated as independent, since there is no direct interaction between each quantum memories. It is likewise considered that $\tau_j=\tau$, $\forall j$, the reliability is written as
\begin{align}
    \mathcal{R}_b(k)
    =&\,\sum_{\bm n,\bm n'}|C(\bm n)|^2|C(\bm n')|^2\cos[\gamma_{\bm{\xi}_0,1}(\tau)\bm 1^{\mathrm{T}}(\bm n-\bm n')]\notag\\
    &\,\times\exp[-(\bm{n}-\bm{n'})^{\mathrm{T}}(\bm{n}-\bm{n'})\Gamma_{\bm{\xi}_0}(\tau)/(2N)],\label{reliability_2_letter}
\end{align}
where $\langle\cdot\rangle_{\Xi_1,\cdots,\Xi_k}$ denotes averaging over all $\Xi_1,\cdots,\Xi_k$, and $C(\bm n)$ is replaced with $\prod_j^k C_{n_j}$. The detailed derivation of the quantum reliability in Eq.~(\ref{reliability_2_letter}) is given in Appendix \ref{subAppendix_Quantum_reliability_deviration}.

We now compare Eqs.~(\ref{reliability_1_letter}) and (\ref{reliability_2_letter}) by setting $C^{(j)}_n=C_n$, $\forall j$ in Eq.~(\ref{reliability_1_letter}), the comparison indicates that the reliability loss of procedure (a) increases significantly with $k$ compared to procedure (b), despite both involving $k$ SIRO processes. Notably, the extra loss arises from the correlations among the collective coupling $\mathcal{G}=\sum_jg_j$ and the collective detuning $\mathcal{D}=\sum_j\Delta_j$ across different SIRO processes. To clarify in detail, we extend to a generic $k$-step SIRO processes, hence the reliability is $\mathcal{R}(k)=\sum_{\bm{n},\bm{n}'}|C(\bm{n})|^2|C(\bm n')|^2\cos[\gamma_{\bm{\xi}_0,1}(\tau)\bm{1}^{\mathrm{T}}(\bm n-\bm n')]\exp[-(\bm n-\bm n')^{\mathrm{T}}\bm\Sigma(\bm n-\bm n')/(2N)]$, where the off-diagonal elements of $\bm{\Sigma}$ are $\bm{\Sigma}_{m\neq l}=\mathrm{Cov}(\mathcal{D}_l,\mathcal{D}_m)/(N\Delta)+\mathrm{Cov}(\mathcal{G}_l,\mathcal{G}_m)/(Ng)$, which represents the correlations between $m,l$-th processes, here $\mathrm{Cov}(\cdot_l,\cdot_m):=\langle(\cdot_l-\langle\cdot_l\rangle)(\cdot_m-\langle\cdot_m\rangle)\rangle$. Since procedure (a) involves repeated operations on the same system, the correlations become maximal, i.e., $\Sigma_{l\neq m}\rightarrow\Sigma_{l=m}=\Gamma_{\bm{\xi}_0}(\tau)$. Therefore, such correlations could significantly reduce the reliability of memory-based devices, whereas enhancing the independence across cycles ($\Sigma_{l\neq m}\rightarrow0$), straightforwardly suppresses the loss. 

Accordingly, when the correlations between different SIRO processes vanish, the reliability $\mathcal{R}(k)$ discussed above attains its upper bound. The upper bound on reliability is Eq.~(\ref{reliability_2_letter}), and it is noteworthy that it can be factored into the reliability of each SIRO process in one memory, i.e., $\mathcal{R}(k)=[\mathcal{R}(1)]^k$, and $\mathcal{R}(1)$ precisely reduces to the state fidelity $\mathcal{R}(1)=\mathcal{F}=\langle|\langle\phi_{\mathrm{out},\bm\xi}|\phi_{\mathrm{in}}\rangle|^2\rangle_{\Xi}$, here
\begin{align}
    \mathcal{F}=&\,\sum_{n,n'}|C_n|^2|C_{n'}|^2\cos[(n-n')\gamma_{\bm{\xi}_0,1}(\tau)]\label{fidelity_letter}\\
    &\,\times\exp\{-(n-n')^2\Gamma_{\bm{\xi}_0}(\tau)/(2N)\}.\notag
\end{align}

The above factorization also indicates that, compared with reliability, $\mathcal{F}$ itself or its product does not provide as direct a characterization of correlation effects among different processes. Physically, this distinction can be understood from the viewpoint of quantum reliability \cite{reliability2023Cui} and the related consistent quantum theory \cite{griffiths1984consistent_quantum_theory,griffiths2003consistent_)quantum_theory}. State fidelity, as well as the corresponding process fidelity and entanglement fidelity, is essentially based on a comparison between the initial and final states. It therefore corresponds to a two-point history, involving only the initial and final events \cite{reliability2023Cui}. By contrast, reliability extends fidelity to the distinction between process trajectories. It not only compares the initial and final events, but also retains the structural information associated with intermediate steps and their correlations, namely a multi-point history. In this sense, it provides a finer characterization of system degradation. A more detailed discussion is given in Appendix \ref{subAppendix_Quantum_reliability}.

To make this point more explicit, we take procedure (a) as an example. As shown in Fig.~\ref{fig2_paper} (a) and in the corresponding trajectory (\ref{trajectory_procedure_a}), we pay additional attention to the events at the intermediate times $\{t_1,t_1+\tau_1,\cdots,t_{k-1}+\tau_{k-1}\}$. As a result, the correlation information between SIRO processes in different time intervals, namely between different input-output two-point histories, also called a two-point trajectory, and the additional influence of these correlations on the device performance can be identified more transparently. This trajectory-level characterization becomes particularly relevant when different process trajectories cannot be treated as independent. For example, in an evolution where environmental degrees of freedom are retained, quantum reliability can extract degradation information beyond the fidelity decay rate through coherence between trajectories \cite{reliability2025Du2}. This indicates that quantum reliability provides a useful trajectory-level perspective for characterizing the degradation of multi-process quantum devices, such as memory-based devices.

It should also be noted that, in the present work, we only introduce additional events at the level of each SIRO process, and thereby reveal the correlation effects across different processes, which may provide guidance for reliable device implementation. A finer-grained characterization, in which more intermediate-time events are included, may extract richer information about device degradation, which deserves further investigation in future work. In the following discussion, apart from the correlations discussed above, $\mathcal{F}$ is sufficient for the main analysis. Therefore, we focus on $\mathcal{F}$ to characterize the degradation of a single SIRO process for convenience.

\subsection{degradation triggered by joint effect between detuning and coupling}\label{Section4_letter}
In this subsection, based on Eq.~(\ref{Gamma_letter}) and Eq.~(\ref{fidelity_letter}), we show that considering coupling disorder alone is insufficient to determine whether it degrades the storage performance. Instead, its effect can be significantly enhanced by its joint action with disordered detuning. This conclusion also indicates that the conventional treatment, in which a single imperfection is analyzed in isolation, needs to be extended to the multi-imperfection case.

There has been some discussion on whether disordered coupling degrades storage performance. For instance, Ref.~\cite{CPSun2003decoherence} conjectured that coupling disorder may induce collective decoherence in the system, but this effect was not demonstrated or elucidated within a concrete storage model. By contrast, Ref.~\cite{X.J.Liu2006Inhomogeneous} focused on disordered coupling as a single imperfection within an explicit storage model and concluded that disordered coupling does not affect the storage performance.

Indeed, it follows from Eq.~(\ref{fidelity_letter}) that in the absence of detuning, one can obtain $\mathcal{F}=1$ (i.e., $\Gamma_{\bm{\xi}_0}=0$), thereby recovering the conclusion of \cite{X.J.Liu2006Inhomogeneous,CPSunAndZSong2005spin,Diniz2011UsingDisorderedCoupling,Lukin2009UsingDisorderedCoupling}. In practice, detuning is inevitable, and the detrimental effect is strongly activated by the presence of detuning, leading to a pronounced reliability loss. This behavior is in sharp contrast to conclusions drawn when disordered coupling is considered in isolation.

Fundamentally, the presence of detuning as an additional adiabatic parameter induces a random Berry's phase in Eq.~(\ref{Berrys phase}), which couples different disorder variables and thereby causing extra degradation. In particular, our results further highlight that some imperfections affect the system primarily through their joint effects, rather than individually.

\section{Refined phase-detuning relation for calibration and Berry-phase-based strategy}\label{Section4_letter}
In this section, based on the actual evolution in Eq.~(\ref{final_state_letter}), we first present a more precise phase-detuning relation for calibrating the global detuning, which refines the conventional empirical relation. Building on this result, we further propose a simple yet effective Berry-phase-based control strategy, which primarily compensates the overall phase shift caused by the global detuning, thereby effectively improving the reliability of quantum memory.

To motivate this discussion, we first note that, in practice, the global detuning $\Delta$ typically arises from multiple competing sources \cite{ACstarkshift2009,ACstarkshift2023,CPSun2004Noabelian_geometry_quantum_memory,Tomography2008squeezed,Tomography2009}, and is therefore often unknown. Eliminating its impact through feasible operations to achieve more reliable quantum memory generally requires an accurate calibration of the global detuning \cite{Tomography2008squeezed,Tomography2009,Tomography2013}. A universal method is to perform quantum tomography \cite{Tomography2008squeezed,Tomography2009,hashimoto2019} on a detection state $|\phi^{(\mathrm{de})}_{\mathrm{in}}\rangle$ that undergoes a SIRO process (before the memory is used for actual storage), since $\Delta$ induces a rotation of Wigner function of an arbitrary input state $|\phi_{\mathrm{in}}\rangle$ around the origin by a mean phase, denoted by $\bar\varphi$, one may infer $\Delta$ from the measured mean phase shift $\bar\varphi$. Based on the actual evolution in Eqs.~(\ref{final_state_letter}) and (\ref{Berry's_phase_expansion_letter}) derived above, a precise phase-detuning relation is obtained,
\begin{equation}
    \bar\varphi=\gamma_{\bm{\xi}_0,1}(\tau)=\Delta(\tau_{\mathrm{s}}+\kappa_\theta\tau_{\mathrm{d}}),\label{tomography_letter}
\end{equation}
Here, to facilitate the subsequent discussion of the driving time $\tau_{\mathrm{d}}$ and storage time $\tau_{\mathrm{s}}$, we define an adiabatic pulse factor $\kappa_\theta:=2\int^{1/2}_0\mathrm{d}(t/\tau_{\mathrm{d}})\sin^2\tilde\theta_{\bm g_0}(t/\tau_{\mathrm{d}})$ with $\tilde\theta_{\bm g_0}(t/\tau_{\mathrm{d}}):=\theta_{\bm g_0}(t)$, so that the driving time $\tau_{\mathrm{d}}$ is explicitly separated out from the Berry's phase. It can be seen that $\kappa_\theta$ is determined solely by the pulse waveform and is independent of the total driving duration, because the time dependence enters only through the dimensionless variable $t/\tau_{\mathrm{d}}$. For example, for the pulse shown in Fig.~\ref{fig1_letter}, $\kappa_\theta\simeq0.29$. The details for the definition of $\kappa_\theta$ is given in Appendix \ref{kappa_zeta}.

We emphasize that achieving accurate calibration for reliable storage requires incorporating the Berry's phase generated by the actual dynamics in current protocols \cite{Tomography2008squeezed,Tomography2009,Tomography2013}, especially the term $\kappa_\theta$ in Eq.~(\ref{tomography_letter}). Otherwise, using the commonly adopted empirical phase-detuning relation, which does not account for this Berry-phase contribution in the actual evolution, can leave a residual detuning
\begin{equation}
\delta^{(\mathrm{de})}=\Delta^{\mathrm{(de)}}-\Delta=\alpha_\theta\tau^{(\mathrm{de})}_\mathrm{d}\Delta/(\tau^{(\mathrm{de})}_\mathrm{s}+\tau^{(\mathrm{de})}_\mathrm{d}/2),\label{residual detuning_letter}
\end{equation}
where $\Delta^{(\mathrm{de})}$ denotes the measured detuning, and $\alpha_\theta:=\kappa_\theta-1/2$.
As a result, even after the measured detuning is used to compensate for the global detuning effect, a phase error $\delta^{(\mathrm{de})}\tau_{s}$ remains in an actual storage process. For a detailed derivation and discussion of this residual detuning, please see Appendix \ref{Section 3_SM}.

Although Eq.~(\ref{residual detuning_letter}) implies the condition $\tau^{(\mathrm{de})}_{\mathrm{s}}/\tau_{\mathrm{d}}^{(\mathrm{de})}\gg 1$ is sufficient to suppress the residue detuning $\delta^{(\mathrm{de})}$, the reliability loss induced by the omission is expected to be significantly amplified by large storage time $\tau_{\mathrm{s}}$ and large photon number fluctuation.

\begin{figure}
    \centering
    \includegraphics[width=1\linewidth]{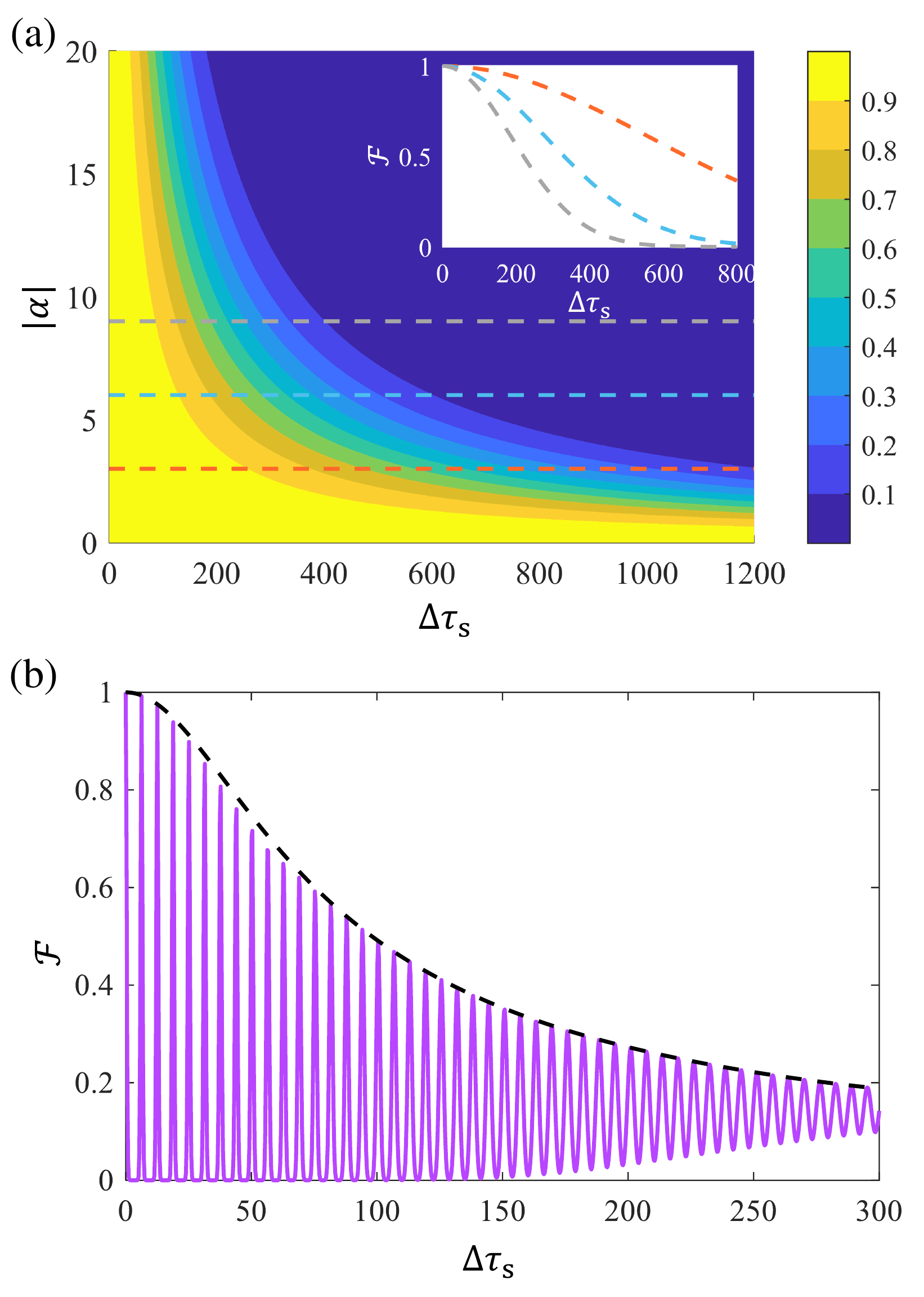}
    \caption{(a) Theoretically predicted fidelity $\mathcal{F}$ of storing a coherent state $|\alpha\rangle$ versus $|\alpha|$ and storage time $\tau_{\mathrm{s}}$. Here, we set $\tau_{\mathrm{s}}^{(\mathrm{de})}/\tau_{\mathrm{d}}^{(\mathrm{de})}=500$ and $\kappa_\theta=0.29$ (thus $\alpha_\theta=-0.21$). Inset represents the fidelity $\mathcal{F}$ versus $\tau_{\mathrm{s}}$ for $|\alpha|=3$ (orange) and $|\alpha|=6$ (blue), $|\alpha|=9$ (gray), which corresponds to the vertical dashed line. (b) Fidelity $\mathcal{F}$ versus $\Delta\tau_{\mathrm{s}}$ with $|\alpha|=2$. The explicit expression of $\mathcal{F}$ is given in Eq.~(\ref{fidelity_coherent_SM}). After applying the Berry-phase-based strategy, the oscillatory solid curve approaches the envelope (dashed curve). Here, we assume $\tau_{\mathrm{s}}\gg\tau_{\mathrm{d}}$ and set $\Delta=10\delta\Delta$.}
    \label{fig3_paper}
\end{figure}

To illustrate this theoretical prediction, an example of storing a coherent state $|\phi_{\mathrm{in}}\rangle=|\alpha\rangle$ is investigated. Let $\delta\Delta=\delta g=0$ for  simplicity. 
After a single SIRO cycle, the fidelity in Eq.~(\ref{fidelity_letter}) is
\begin{equation}
\mathcal{F}\simeq\exp[-2|\alpha|^2(1-\cos(\delta^{(\mathrm{de})}\tau_{\mathrm{s}}))].\label{fidelity_coherent_state}
\end{equation}
The theoretical curve in Fig.~\ref{fig3_paper} (a) shows the fidelity (\ref{fidelity_coherent_state}) drops significantly with increasing $|\alpha|$ and $\tau_{\mathrm{s}}$. More generally, by expanding Eq.~(\ref{fidelity_letter}) to leading term, we obtain $\tau_\mathrm{s}^{(\mathrm{de})}/\tau_\mathrm{d}^{(\mathrm{de})}\gg\Delta \tau_s\langle\Delta n^2\rangle^{1/2}$ to ensure high fidelity. It shows achieving a universal quantum memory with long storage time requires more constraints on $\tau_{\mathrm{s}}^{(\mathrm{de})}$ and $\langle\Delta n^2\rangle:=\langle (\hat a^\dagger \hat a)^2\rangle-\langle\hat a^\dagger \hat a\rangle^2$ of stored states.
Therefore, Eq.~(\ref{tomography_letter}) could ease practical calibration, directly supporting long-time and high-capacity storage.

Building on the precise relation in Eq.~(\ref{tomography_letter}) for calibration, we further propose a simple yet powerful Berry-phase-based strategy to compensate the overall phase shift caused by the global detuning, thereby effectively improving the storage performance.

More specifically, once $\Delta$ is accurately calibrated, it is indicated by Eq.~(\ref{tomography_letter}) that enforcing $\gamma_{\bm{\xi}_0,1}(\tau)\rightarrow2n\pi$ effectively cancels the overall phase shift. In this case, one can obtain $\cos[(n-n')\gamma_{\bm{\xi}_0,1}(\tau)]\rightarrow1$ in Eq.~(\ref{fidelity_letter}). To illustrate the resulting mitigation, we continue to use the coherent state $|\phi_{\mathrm{in}}\rangle=|\alpha\rangle$ as an example. As shown in Fig.~\ref{fig3_paper} (b), imposing this condition markedly enhances $\mathcal{F}$ as a function of the storage time $\tau_{\mathrm{s}}$, lifting the performance from the oscillating solid curve toward the envelope (dashed curve). 

Physically, the effective strategy $\gamma_{\bm{\xi}_0,1}(\tau)\rightarrow2n\pi$ can be implemented in several complementary ways. A direct approach is to tune the optical frequency. In protocols where detuning is intentionally introduced to improve efficiency \cite{Tomography2008squeezed} or to realize geometric quantum memory \cite{CPSun2004Noabelian_geometry_quantum_memory}, the same target can instead be reached by requiring the driving time to satisfy $\kappa_{\theta}\tau_{\mathrm{d}}=2n\pi/\Delta-\tau_{\mathrm{s}}$, or by applying an additional compensation gate $U_{\mathrm{com}}=\exp(-\mathrm{i}\gamma_{\bm{\xi}_0,1}(\tau)\hat a^\dagger \hat a)$ to $|\phi_{\mathrm{out}}\rangle$.

It should be emphasized that although the strategy introduced above can effectively improve the memory performance by compensating the global phase shift, disorder-induced degradation still persists. It is therefore necessary to investigate the residual effect induced by disorders in the next section.

\section{Effects of disorders}\label{Section5_letter}
In this section, we further investigate the residual disorder effects on the memory arising from inhomogeneous broadening $\delta\Delta$ and disordered coupling $\delta g$. Following the discussion of the strategy in Sec.~\ref{Section4_letter}, we take $\cos[(n-n')\gamma_{\bm{\xi}_0,1}(\tau)]\rightarrow1$. We first show that, for a broad class of stored states relevant to practical applications and satisfying two conditions, disorders constrain the stored state solely through its photon number fluctuation $\langle\Delta n^2\rangle$ in the regime $\langle\Delta n^2\rangle\gg1$ when fidelity is fixed. This constraint follows a universal inverse proportionality. We then illustrate this result with representative examples and further reveal other universal behaviors. Finally, we demonstrate a universal lower bound on the fidelity for an arbitrary stored state, which likewise depends solely on $\langle\Delta n^2\rangle$. These results provide the basis for the universal capacity-time trade-off derived in the next section.

Taking physical realizability into account, we focus here on a class of stored states that satisfy two conditions. This class includes several significant states in quantum communication and quantum computation, such as cat states and squeezed states \cite{CAT2013,GKP2024,Tomography2008squeezed}. The two conditions are as follows: (1) $|C_n|^2$ of the states $|\phi_{\mathrm{in}}\rangle=\sum_n C_n |n\rangle_L$ is well localized around the mean photon number $\langle n\rangle$; and (2) in the regime $\langle\Delta n^2\rangle\gg1$, the states exhibit a stable normalized fluctuation distribution $Z_{n,\langle\Delta n^2\rangle}:=(n-\langle n\rangle)/\langle\Delta n^2\rangle^{1/2}$, namely, the profile of $Z_{n,\langle\Delta n^2\rangle}$ is insensitive to variations in $\langle\Delta n^2\rangle$ and $\langle n\rangle$. It should be noted that these two conditions are formulated in mathematical terms in the Appendix \ref{condition_mathematic}, where their formulations—including the degree of localization and the stability of $Z_{n,\langle\Delta n^2\rangle}$—are given explicitly. The following analysis relies on these formulations.

As a qualitative interpretation of the second condition, the corresponding statistical description remains robust under nonideal perturbations, and is therefore amenable to experimental use. This is because, when $\langle\Delta n^2\rangle\gg1$, although nonideal effects may modify the overall characteristics of a state, such as $\langle n\rangle$ and $\langle\Delta n^2\rangle$, the normalized fluctuation distribution remains approximately unchanged to some extent. As a result, states obtained in different experimental runs can still be compared on the basis of the same statistical distribution $Z_{n,\langle\Delta n^2\rangle}$.

With these conditions in place, we now present the corresponding universal results. Under the two conditions, we find that in the regime of $\langle\Delta n^2\rangle\gg1$, the fidelity can be approximated as a universal form
\begin{equation}
    \mathcal{F}\simeq\sum_m\frac{c_m}{m!}[-\Gamma_{\bm{\xi}_0}(\tau)\langle\Delta n^2\rangle/N]^m,\label{fidelity_large_fluctuation_letter}
\end{equation}
where $c_m$ is determined by the stored state and index $m$. The detailed discussion and proof are provided in Appendix \ref{Section 4_SM}. It can be observed from Eq.~(\ref{fidelity_large_fluctuation_letter}) that $\Gamma_{\bm{\xi}_0}(\tau)$ (i.e., detuning and disordered coupling) constrains the stored state solely through $\langle\Delta n^2\rangle$, with a universal inverse proportionality for fixed fidelity.

\begin{figure}
    \centering
    \includegraphics[width=1\linewidth]{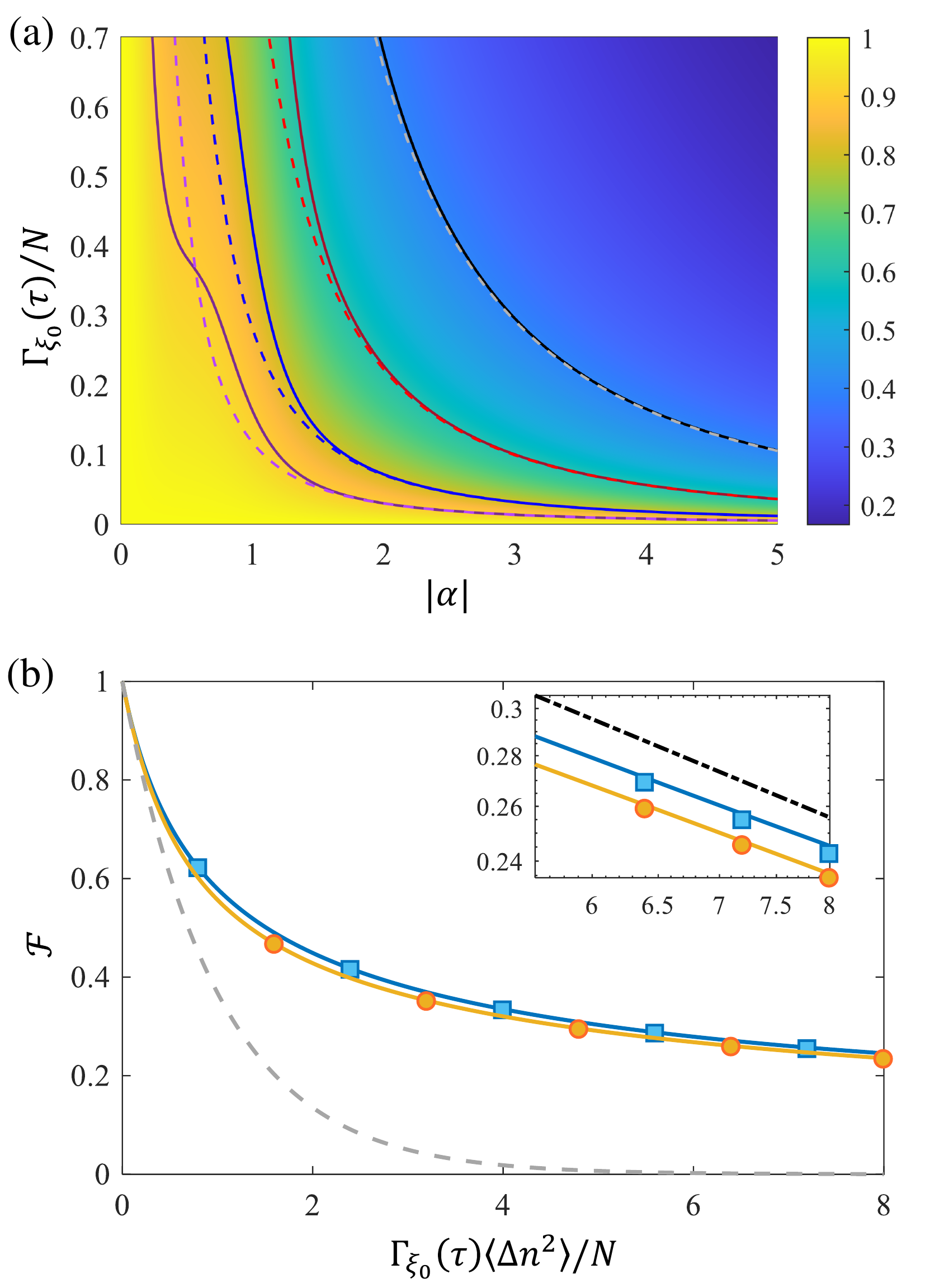}
    \caption{(a) Fidelity $\mathcal{F}$ versus $\Gamma_{\bm{\xi}_0}(\tau)/N$ and $|\alpha|$, when storing cat state $\mathcal{N}_0(|\alpha\rangle+\exp(\eta+\mathrm{i}\theta)|-\alpha\rangle)$, $\eta,\ \theta\in\mathbb{R}$, here we set $\eta=1,\ \theta=\pi$. The curves represent contours of $\mathcal{F}=0.9,\ 0.8,\ 0.6,\ 0.4$ from left to right, with solid lines for the exact form, and dashed lines for the approximate form when $\langle\Delta n^2\rangle\gg1$ (here $\langle\Delta n^2\rangle\simeq|\alpha|^2$), following a relation $\Gamma_{\bm{\xi_0}}(\tau)/N\propto1/|\alpha|^2$. (b) $\mathcal{F}$ versus $\Gamma_{\bm{\xi}_0}\langle\Delta n^2\rangle/N$. The blue and orange solid lines represent storing the cat state ($\eta=0,\ \theta=0$) and the uniform superposition state, respectively, with $\langle\Delta n^2\rangle=10$. Squares and circles represent the approximate form. The dashed line is lower bound for arbitrary state fidelity, and the inset shows scaling as $(\Gamma_{\bm{\xi}_0}(\tau)\langle\Delta n^2\rangle/N)^{-1/2}$. Please see Appendix \ref{Section 5_SM} for detailed derivation and such forms of fidelity.}
    \label{fig4_paper}
\end{figure}
To illustrate this universal relation, cat states and uniform superposition state are used as storage examples. The cat states are taken in the form $\mathcal{N}_0(|\alpha\rangle_L+\exp(\mathrm{\eta+\mathrm{i}\theta})|-\alpha\rangle_L)$, $\forall\ \eta,\,\theta\in\mathbb{R}$, while the uniform superposition state is given by $\sum_{n}|n\rangle_L/\sqrt{N_c}$. Fig.~\ref{fig4_paper} (a) shows the exact form of fidelity for cat state (please see Eq.~(\ref{fidelity_cat_SM}) in Appendix \ref{Section 5_SM}) versus $\langle\Delta n^2\rangle$ and $\Gamma_{\bm{\xi}_0}(\tau)/N$, it can be observed that the constant-fidelity contours (solid lines) converge towards the hyperbolic dashed line which represents the inverse proportionality $\Gamma_{\bm{\xi}_0}(\tau)/N\propto1/\langle\Delta n^2\rangle$. 

Moreover, Fig.~\ref{fig4_paper} (b) shows the behaviors of the exact fidelity of these states as a function of $\Gamma_{\bm{\xi}_0}(\tau)\langle\Delta n^2\rangle/N$, the squares and circles denote the approximate fidelity in large $\langle\Delta n^2\rangle$ limits, consistent with Eq.~(\ref{fidelity_large_fluctuation_letter}). More specifically, for cat state (circles), the approximate fidelity is $\mathcal{F}_C\simeq \{1+2\langle\Delta n^2\rangle\Gamma_{\bm{\xi}_0}(\tau)/(2N)\}^{-1/2}$, while for uniform superposition state (squares), it reads
\begin{equation}
    \mathcal{F}_{U}\simeq\sqrt{\pi}\mathrm{erf}(x)/x-\{\exp(-x^2)-1\}/x^2,
\end{equation}
with $x= 6\langle\Delta n^2\rangle\Gamma_{\bm{\xi}_0}(\tau)/(2N)$. Notably, it is observed from inset that both forms exhibit power-law tail behavior, scaling as $(\Gamma_{\bm{\xi}_0}(\tau)\langle\Delta n^2\rangle/N)^{-1/2}$. The detailed derivation and behaviors of fidelity are provided in Appendix \ref{Section 5_SM}. 

More generally, it can be proven that (please see Appendix \ref{Section 6_SM}) the fidelity in Eq.~(\ref{fidelity_letter}) satisfies
\begin{equation}
    \mathcal{F}\geq\exp(-\Gamma_{\bm{\xi}_0}(\tau)\langle\Delta n^2\rangle/N),\label{lowe_bound_letter}
\end{equation}
for an arbitrary stored state and $\langle\Delta n^2\rangle$, where the lower bound (gray dashed line in Fig.~\ref{fig4_paper} (a)) exhibits universal exponential decay and provides a general benchmark for the residual disorder-induced degradation.

These universal results above suggest that the induced disorder effects may impose a limitation on the information-storage capacity $\mathcal{C}$. Qualitatively, when $\langle\Delta n^2\rangle$ of reliably storable states is constrained, the effective region that such states can stably occupy in the Fock basis may also become restricted. As a consequence, the amount of quantum information that the memory can carry may be correspondingly reduced. These results therefore provide the basis for presenting a universal trade-off between capacity and storage time in the next section.

\section{trade-off in a reliable universal quantum memory}\label{Section6_letter}
In this section, based on the results of Sec.~\ref{Section5_letter}, we present a universal capacity-time trade-off in a highly reliable universal quantum memory. This trade-off provides guidance for the realization of large information-storage capacity universal quantum memory in the presence of decoherence induced by static inhomogeneous broadening $\delta\Delta$ and disordered coupling $\delta g$. We also provide the adiabatic conditions for a reliable SIRO process, since the EIT-based quantum memory relies on adiabatic manipulation. Finally, taking a cold-atom platform as an example, we give a brief and intuitive order-of-magnitude illustration of the trade-off.

Drawing from Sec.~\ref{Section5_letter}, a general trade-off for storage capacity, storage and driving time of a quantum memory can be presented as 
\begin{equation}
    1-\mathcal{F}_0\simeq(\delta\Delta^2\tau_{\mathrm{s}}^2+\Delta^2\delta g^2\zeta_\theta^2\tau_{\mathrm{d}}^2/g^2)(\exp\mathcal{C}-1)^2/(4N),\label{trade_off_letter}
\end{equation}
in the highly reliable region. Notably, the effective information-storage capacity is found to be directly related to the photon number fluctuation, reads
\begin{equation}
    \mathcal{C}\simeq\ln\{(4\langle\Delta n^2\rangle_{\max})^{1/2}+1\},\label{capacity_form_letter}
\end{equation}
where the derivation and details of Eqs.~(\ref{trade_off_letter}) and (\ref{capacity_form_letter}) are presented in Appendix \ref{Section 7_SM}, here $\langle\Delta n^2\rangle_{\max}$ denotes the maximum photon number fluctuation allowed for stored states, $\mathcal{F}_0$ is the corresponding fidelity, while also serving as the lower bound for all stored states supported by this capacity. Here, analogous to the definition of $\kappa_\theta$, we introduce another adiabatic pulse factor $\zeta_\theta:=\int^{1/2}_0\mathrm{d}(t/\tau_{\mathrm{d}})\sin^2(2\tilde{\theta}_{\bm g_0}(t/\tau_{\mathrm{d}}))$. The details is given in Appendix \ref{kappa_zeta}. For the pulse waveform shown in Fig.~\ref{fig1_letter}, we have $\zeta_\theta\simeq0.05$.

It is observed that the broadening $\delta\Delta$ primarily constrains the storage time $\tau_{\mathrm{s}}$, but the disordered coupling $\delta g$ and global detuning $\Delta$ solely limits the driving time $\tau_{\mathrm{d}}$. Moreover, there exists a constraint 
\begin{equation}
    \tau_{\mathrm{d}},1/|\Delta|\gg\sqrt{n_{\mathrm{max}}/(Ng^2)},
\end{equation}
with $\sum_{n_{\max}}^{\infty}|C_n|^2\ll1$ based on adiabatic condition (please see Appendix \ref{Section 2_SM} for derivation). Therefore, these results provide precise targets for optimizing the relevant quantities.

To further illustrate the trade-off, the recently demonstrated cold-atom platform for quantum memory \cite{CJF2024cold_atom,GGC2014cold_atom,Sayrin2015fiber1,RZhao2009long_lived_quantum_memory,SLZhu2019SinglePhotonState} is considered. As an order-of-magnitude estimate, the atomic cloud width is $d\sim1$$\mathrm{cm}$ \cite{SLZhu2019SinglePhotonState}, the signal beam waist is $w_0\sim 100\mu \mathrm{m}$ and the typical atomic density is $n\sim 10^{11}\mathrm{cm}^{-3}$ \cite{Sayrin2015fiber1}, leading to the effective atom number of $N\sim nw_0^2d\sim10^7$. For a broadening $\delta\Delta\sim1\mathrm{MHz}$ in EIT-memory \cite{Philipp2022singlePhotonState}, achieving $\mathcal{F}_0=0.9$ requires time and capacity to satisfy the trade-off $\tau_{\mathrm{s}}\exp\mathcal{C}\simeq1$ms, which likewise requires all stored states to obey $\langle\Delta n^2\rangle^{1/2}\tau_{\mathrm{s}}\lesssim1\mathrm{ms}$. 

It is noteworthy that many protocols deliberately introduce detuning or disordered coupling to enhance storage efficiency \cite{Tomography2008squeezed}, to realize geometric quantum memory \cite{CPSun2004Noabelian_geometry_quantum_memory}, or to effectively suppress decoherence by exploiting nuclear spin waves \cite{CPSunAndZSong2005spin,Lukin2009UsingDisorderedCoupling}. In addition, AC-stark shifts from strong control field can reach MHz scale \cite{HYF2018ac_stark_shift_storage_efficiency}. These situations underscore the need to carefully address the correlated effects of detuning and disordered coupling. As an illustration, with average coupling strength $g\sim 0.1\mathrm{MHz}$ and $\delta g/g\sim 0.1$, there exists a trade-off $\Delta\tau_{\mathrm{d}}\exp\mathcal{C}\simeq10^5$ (equivalently $\Delta\tau_{\mathrm{d}}\langle\Delta n^2\rangle^{1/2}\lesssim10^5$ for stored states) to maintain $\mathcal{F}_0=0.9$, and the adiabatic condition demands $|\Delta|,1/\tau_{\mathrm{d}}\ll10^{2.5}/n^{1/2}_{\max}$MHz $\sim10^{2.5}/\langle\Delta n^2\rangle^{1/2}$MHz. Notably, recent proposals for fiber-coupled quantum memory \cite{Sayrin2015fiber1,gouraud2015fiber2,Leong2020fiber3,Blatt2016fiber4,Xin2019fiber5}, which guide the retrieval light efficiently, could reduce the effective atom number to $N\sim10^3-10^4$ \cite{gouraud2015fiber2,Leong2020fiber3,Xin2019fiber5}, thereby making the trade-off become more stringent.

\section{conclusion}\label{Section7_letter}
In this work we investigate the reliability and performance limits of an EIT-based quantum memory in the presence of static disordered coupling and detuning. Through a unified approach, we show that these disorders jointly induce a random Berry's phase, which degrades the performance of the memory. We first focus on the overall phase shift caused by the global detuning, which is often induced in practice by effects such as the AC-stark shift. We show that the resulting dephasing can be mitigated through an optimal control strategy enabled by accurate detuning calibration. Within a comprehensive model, we provide a refined phase–detuning relation for global-detuning calibration \cite{Tomography2009,Tomography2008squeezed,Tomography2013}, reducing the systematic errors inherent in commonly used relations, particularly over long storage times. Building on this result, we introduce a Berry-phase-based control scheme to compensates the overall phase shift caused by the global detuning and effectively improves the performance of the quantum memory.

Residual disorder effects that remain beyond the scheme are further investigated. A universal constraint on the photon-number fluctuation $\langle\Delta n^2\rangle$ is obtained for a broad class of states relevant to practical applications. Based on the universal result, and driven by the requirement of universality of the quantum memory, an effective measure of information-storage capacity quantifying the supported information content is introduced, and an intrinsic trade-off among storage capacity, storage time, and control-pulse driving time is established. These results provide a quantitative basis for optimizing reliable operation of universal quantum memories.

Our unified treatment further reveals a new phenomenon: multiple disorder sources act cooperatively to affect the quantum memory through complex joint effects, rather than through independent contributions. Such joint effects can tighten the driving time of control pulse and lead to behavior that differs from conclusions drawn from commonly used one-factor analyses. This perspective is directly relevant to the experimental protocols involving disordered coupling \cite{CPSunAndZSong2005spin,Diniz2011UsingDisorderedCoupling} and detuning \cite{CPSun2004Noabelian_geometry_quantum_memory,Tomography2008squeezed}, and it could also be useful for geometric and holographic quantum computation \cite{ZHANG2023geometric_QuantumComputation,Zanardi2005geometricQuantumComputation}. This phenomenon also indicates the need to move beyond the conventional treatment in which different imperfections are analyzed in isolation.

Finally, based on quantum reliability as a process-oriented measure, the reliability of typical processes in memory-based devices \cite{repeaters2019Pirandola,repeaters2023Azuma,synchronization2007} is examined. The results suggest that reducing correlations of disorder across different storage processes can effectively suppress overall reliability loss at the device level. This shows that, in addition to optimizing a single storage process, device-level optimization of the correlations between different processes is also an important route toward highly reliable quantum devices.

It should be further emphasized that, although this work provides a unified treatment and reveals the joint effect induced by detuning and coupling disorder, together with corresponding mitigation strategies, many other imperfections still hinder the reliable implementation of quantum memory in current experiments. Examples include atomic thermal motion \cite{fleischhauer2002quantum_memoryPRA}, photon loss \cite{lvovsky2009optical_quantum_memory}, background gas collisions \cite{Bali1999background_collision,Manz2007background_collision}, and other loss mechanisms \cite{Four-wave-mixing2013,Four-wave-mixing2014,Four-wave-mixing2019,Spin-wave2015PRL,Spin-wave2019PRL}. These effects are also important sources that limit the performance and lifetime of quantum memories, as well as the realization of memory-based devices. From this perspective, the unified treatment of different nonideal factors developed here, together with the advantages of quantum reliability as a measure, should be extended in future work to include more realistic factors. Such extensions may reveal new joint effects among different nonideal sources and provide theoretical guidance for designing new mitigation strategies.

\section*{Acknowledgment}
This work was supported by the Science Challenge Project (Grant No. TZ2025017), the National Natural Science Foundation of China (NSFC) (Grant No. 12088101), the NSAF (Grant No. U2330401).

\section*{Data Availability}
The data that support the findings of this article are openly available \cite{Data_availability}.

\appendix

\begin{widetext}

\section{Eigen-spectrum Derivation in the Large-atom numbers and Low-excitation Limits}\label{Section 1_SM}
Under the consideration of $\delta_j\equiv0$ and $\Omega_j(t)\equiv\Omega(t)$, $\forall j$-th atom, the Hamiltonian in Eq.~(\ref{Hamiltonian_initial_letter}) for a specific set of $\bm{\Delta}=(\Delta_1,\cdots,\Delta_N)$ and $\bm{g}=(g_1,\cdots,g_N)$ reads
\begin{equation}
    \hat H=\hat{a}\sum_{j=1}^N g_j\exp(\mathrm{i}\Delta_j t)\exp(\mathrm{i}\bm{K}_{ba}\cdot\bm{r}_j)\hat\sigma_{ab}^{(j)}+\Omega(t)\sum_{j=1}^N\exp(\mathrm{i}\bm{K}_{ca}\cdot\bm{r}_{j})\hat\sigma_{ac}^{(j)}+\mathrm{H.c.}.\label{Hamiltonian_initial_SM}
\end{equation}
We define the quasi-spin-wave collective excited operator 
\begin{equation}
    \hat S_{\bm{\xi}}=\sum_j^N g_j\hat \sigma_{bs}^{(j)}\exp(-\mathrm{i}\bm{K}_{bs}\cdot\bm{r}_j-\mathrm{i}\Delta_jt)/(\sum_k^N g_k^2)^{1/2},\label{collective_excited_operator_SM}
\end{equation}
with $\hat S_{\bm{\xi}}=\hat A_{\bm{\xi}},\ \hat C_{\bm{\xi}}$ for $s=a,\ c$, the subscript $\bm{\xi}=(\bm{\Delta},\bm{g})$ denotes the set of $\bm{\Delta}$ and $\bm{g}$, the wave vector satisfying $\bm{K}_{bc}=\bm{K}_{ba}-\bm{K}_{ca}$ for second transition process from $|b\rangle\rightarrow|c\rangle$.

With $N\gg1$ and low excitation condition $\langle\hat A^\dagger_{\bm{\xi}} \hat A_{\bm{\xi}}\rangle,\ \langle \hat C^\dagger_{\bm{\xi}} \hat C_{\bm{\xi}}\rangle \ll 1$, i.e., there are negligible number of occupation numbers in state $|a\rangle$ and $|c\rangle$ compared to $N$, the spin-wave operators can be treated as bosons, obeying the commutation relation $[\hat S_{\bm{\xi}},\hat S^\dagger_{\bm{\xi}}]\simeq1$. Furthermore, these two operators are independent of each other with $[\hat A_{\bm{\xi}},\hat C_{\bm{\xi}}]=0, [\hat A_{\bm{\xi}},\hat C^\dagger_{\bm{\xi}}]=-\sum_j^N g_j^2 \hat \sigma_{ca}^{(j)}\exp(-\mathrm{i}\bm{K}_{ca}\cdot\bm{r}_j)/\sum_k g_k^2\simeq0$.

Under the conditions, the Hamiltonian in Eq.~(\ref{Hamiltonian_initial_SM}) is rewritten in a compact form as
\begin{equation}
\hat H=\varepsilon(\hat a\hat A^\dagger_{\bm{\xi}}\sin\theta_{\bm{g}}+\hat T_+\cos\theta_{\bm{g}}+\mathrm{H.c.}),\label{Hamiltonian_fin_SM}
\end{equation}
where $\varepsilon=(\sum_jg_j^2+\Omega^2(t))^{1/2}$ and $\tan\theta_{\bm g}=(\sum_jg_j^2)^{1/2}/\Omega(t)$. The quasi-angular-momentum operator denotes $\hat T_-=\sum_j \exp(-\mathrm{i}\mathbf{K}_{ca}\cdot\mathbf{r}_j)\hat \sigma_{ca}^{(j)}, \hat T_+=(\hat T_-)^\dagger$, and we define $\hat T_3:=\sum_j(\hat \sigma_{aa}^{(j)}-\hat \sigma_{cc}^{(j)})/2$. Here, $\hat T_+$, $\hat T_-$ and $\hat T_3$ generate $\mathrm{SU}(2)$ algebra with commutation relations $[\hat T_3,\hat T_\pm]=\pm \hat T_\pm, [\hat T_+,\hat T_-]=2\hat T_3$. 

Furthermore, we define the Heisenberg–Weyl algebra $h_{\bm{\xi}}=\{\hat A_{\bm{\xi}},\hat A^\dagger_{\bm{\xi}},\hat C_{\bm{\xi}},\hat C^\dagger_{\bm{\xi}},1\}$ for the atomic collective excitations. We obtain that
\begin{align}
    [\hat T_-,\hat C_{\bm{\xi}}]=&\,-\hat A_{\bm{\xi}},\label{algebra relation1_SM}\\
    [\hat T_+,\hat A_{\bm{\xi}}]=&\,-\hat C_{\bm{\xi}}\label{algebra relation2_SM},
\end{align}
which indicates $h_{\bm{\xi}}$ and $\mathrm{SU}(2)=\{\hat T_+,\hat T_-,\hat T_3\}$ satisfy semi-direct product algebraic relation as $[\mathrm{SU}(2),h_{\bm{\xi}}]\subset h_{\bm{\xi}}$. Therefore, the Hamiltonian in Eq.~(\ref{Hamiltonian_fin_SM}) can be depicted by a dynamic Lie algebra $\mathrm{SU}(2)\overline{\bigotimes}h_{\bm{\xi}}\bigotimes h_{\hat a}$, where the Heisenberg-Weyl algebra $h_{\hat a}:=\{\hat a,\hat a^\dagger,1\}$ represents the optical field.

Based on the symmetry of the system, we can generate all eigen-spectrum via the standard dynamic algebra method \cite{wybourne1974DynamicAlgebra,CPSun2003quasi}. The ground state of system is denotes as $|\bm{0}\rangle=|0\rangle_L\otimes|0\rangle_A$ with $H|\bm{0}\rangle=0$, where $|0\rangle_L$ denotes the vacuum of quantum optical field and $|0\rangle_A:=\prod_j|b\rangle_j$ is the ground state of the atomic ensemble. Based on the algebra relations in Eqs.~(\ref{algebra relation1_SM}) and (\ref{algebra relation2_SM}), we can straightforwardly construct a Schwinger representation $\hat T_-=\hat A_{\bm{\xi}}\hat C^\dagger_{\bm{\xi}}$ and $\hat T_+=\hat C_{\bm{\xi}}\hat A^\dagger_{\bm{\xi}}$, to map the SU(2) algebra onto the bosonic $\hat A_{\bm{\xi}}$, $\hat C_{\bm{\xi}}$ space. Hence, the Hamiltonian in Eq.~(\ref{Hamiltonian_fin_SM}) reads
\begin{equation}
    \hat H=\varepsilon(\hat A_{\bm{\xi}}\hat B^\dagger_{\bm{\xi}}+\mathrm{H.c.}).\label{Hamiltonian_bosonic_SM}
\end{equation}
Here we can define the bright (dark)-state operator
\begin{align}
    \hat B_{\bm{\xi}}=&\,\hat a\sin\theta_{\bm{g}}+\hat C_{\bm{\xi}}\cos\theta_{\bm{g}},\\
    \hat D_{\bm{\xi}}=&\,\hat a\cos\theta_{\bm{g}}-\hat C_{\bm{\xi}}\sin\theta_{\bm{g}},\label{dark_operator_SM}
\end{align}
which also satisfy the bosonic commutation relations $[\hat D_{\bm{\xi}},\hat D_{\bm{\xi}}^\dagger]=1$, $[\hat B_{\bm{\xi}},\hat B_{\bm{\xi}}^\dagger]=1$. Notably, the mode $\hat D_{\bm{\xi}}$ obeys $[\hat H,\hat D_{\bm{\xi}}]=0$, here we can obtain series of eigen-states with vanishing eingen-value, reffed to as dark-states
\begin{equation}
    |D_{\bm{\xi},n}\rangle=\frac{1}{\sqrt{n!}}(\hat D^\dagger_{\bm{\xi}})^n|\mathbf{0}\rangle.
\end{equation}

We further diagonalize the Hamiltonian in Eq.~(\ref{Hamiltonian_bosonic_SM}) to construct all eigen-states, yielding
\begin{equation}
   \hat H=\varepsilon(\hat Q_{\bm{\xi},+}^\dagger \hat Q_{\bm{\xi},+}-\hat Q_{\bm{\xi},-}^\dagger \hat Q_{\bm{\xi},-}),
\end{equation}
where
\begin{equation}
    \hat Q_{\bm{\xi},\pm}=\frac{1}{\sqrt{2}}(\hat A_{\bm{\xi}}\pm \hat B_{\bm{\xi}}),\label{Q_SM}
\end{equation}
and $[\hat Q_{\bm{\xi},i},\hat Q^\dagger_{\bm{\xi},j}]=\delta_{ij}$, $i,j=+,-$. Hence eigen-states in $N\gg1$ and low excitation condition are expresses as
\begin{equation}
    |e_{\bm{\xi}}(m,k,n)\rangle=\frac{1}{\sqrt{m!k!n!}}(\hat Q_{\bm{\xi},+}^\dagger)^m (\hat Q_{\bm{\xi},-}^\dagger)^k(\hat D^\dagger_{\bm{\xi}})^n|\mathbf{0}\rangle.
\end{equation}
with eigenvalue $e_{\bm{\xi}}(m,k,n)=(m-k)\varepsilon$determined by a family of $good$ quantum number $m,k,n$, and $|D_{\bm{\xi},n}\rangle\equiv|e_{\bm{\xi}}(0,0,n)\rangle$

\section{derivation of Adiabatic condition, High-order adiabatic condition and Berry's phase}\label{Section 2_SM}
Before deriving the adiabatic condition, Berry's phase and high-order adiabatic condition, we present some preliminary results.

(1) Base on Eq.~(\ref{dark_operator_SM}) and Eq.~(\ref{Q_SM}), we can derive that
\begin{align}
    \partial_{\theta_{\bm{g}}} \hat D_{\bm{\xi}}^\dagger=&\,-\frac{1}{\sqrt{2}}(\hat Q_{\bm{\xi},+}^\dagger-\hat Q_{\bm{\xi},-}^\dagger),\\
    \partial_{\theta_{\bm{g}}} \hat Q_{\bm{\xi},\pm}^\dagger=&\,\pm \frac{1}{\sqrt{2}} \hat D^{\dagger}_{\bm{\xi}}.
\end{align}
Therefore,
\begin{align}
    \langle e_{\bm{\xi}}(m',k',n')|\partial_{\theta_{\bm{g}}}|e_{\bm{\xi}}(m,k,n)\rangle=&\,\langle e_{\bm{\xi}}(m',k',n')|\{\sqrt{m}(\partial_{\theta_{\bm{g}}}\hat Q_{\bm{\xi},+}^\dagger)|e_{\bm{\xi}}(m-1,k,n)\rangle+
    \sqrt{k}(\partial_{\theta_{\bm{g}}}\hat Q_{\bm{\xi},-}^\dagger)|e_{\bm{\xi}}(m,k-1,n)\rangle\notag\\
    &\,+\sqrt{n}(\partial_{\theta_{\bm{g}}}\hat D_{\bm{\xi}}^\dagger)|e_{\bm{\xi}}(m,k,n-1)\rangle\}\notag\\
    =&\,\sqrt{\frac{m}{2}}\langle e_{\bm{\xi}}(m',k',n')|\hat D_{\bm{\xi}}^\dagger|e_{\bm{\xi}}(m-1,k,n)\rangle-\sqrt{\frac{k}{2}}\langle e_{\bm{\xi}}(m',k',n')|\hat D_{\bm{\xi}}^\dagger|e_{\bm{\xi}}(m,k-1,n)\rangle\notag\\
    &\,-\sqrt{\frac{n}{2}}\langle e_{\bm{\xi}}(m',k',n')|(\hat Q_{\bm{\xi},+}^\dagger-\hat Q_{\bm{\xi},-}^\dagger)|e_{\bm{\xi}}(m,k,n-1)\rangle\label{partial_theta_adiabatic_SM}\\
    =&\,\sqrt{\frac{n+1}{2}}\delta_{n',n+1}(\sqrt{m}\delta_{m',m-1}\delta_{k',k}-\sqrt{k}\delta_{m',m}\delta_{k',k-1})-\sqrt{\frac{n}{2}}(\sqrt{m+1}\delta_{m',m+1}\delta_{k',k}\notag\\
    &\,-\sqrt{k+1}\delta_{m',m}\delta_{k',k+1})\delta_{n',n-1}\notag.
\end{align}

(2) By defining the adiabatic variable $\phi_j(t)=\Delta_j t$ for $j$-th atom, we have
\begin{align}
    \partial_{\phi_j} \hat D_{\bm{\xi}}^\dagger=&\,-\sin\theta_{\bm{g}}\partial_{\phi_j} \hat C_{\bm{\xi}}^\dagger,\\
    \partial_{\phi_j} \hat Q_{\bm{\xi},\pm}^\dagger=&\,\frac{1}{\sqrt{2}}(\partial_{\phi_j} \hat A_{\bm{\xi}}^\dagger\pm\cos\theta_{\bm{g}}\partial_{\phi_j} \hat C_{\bm{\xi}}^\dagger),
\end{align}
with
\begin{equation}
    \partial_{\phi_j}\hat S^\dagger_{\bm{\xi}}=\mathrm{i}g_j\exp(\mathrm{i}\phi_j+\mathrm{i}\bm{K}_{bs}\cdot\bm r_j)\hat \sigma_{sb}^{(j)}/(\sum_k^N g_k^2)^{1/2}.\label{partial_phi_collective_excited_operator_SM}
\end{equation}
Furthermore, We derive that $[\hat S'^{\dagger}_{\bm{\xi}},\partial_{\phi_j} \hat S^\dagger_{\bm{\xi}}]=0$, 
with $\hat S_{\bm{\xi}},\hat S'_{\bm{\xi}}=\hat A_{\bm{\xi}}, \hat C_{\bm{\xi}}$ for $s,s'=a,c$. Based on this, we obtain
\begin{equation}
    \partial_{\phi_j}(\hat D^\dagger_{\bm{\xi}})^n=\sum_l(\hat D^\dagger_{\bm{\xi}})^l(\partial_{\phi_j}\hat D^\dagger_{\bm{\xi}})(\hat D^\dagger_{\bm{\xi}})^{n-1-l}=n(\hat D^\dagger_{\bm{\xi}})^{n-1}\partial_{\phi_j}\hat D^\dagger_{\bm{\xi}},\label{partial_phi_Dn_SM}
\end{equation}
as well as $\partial_{\phi_j} (\hat Q_{\bm{\xi},\pm}^\dagger)^n=n(\hat Q_{\bm{\xi},\pm}^\dagger)^{n-1}\partial_{\phi_j} \hat Q_{\bm{\xi},\pm}^\dagger$, $\forall n$.

(3) We define the un-normalized state as
\begin{equation}
    |\tilde{1}_{s_j}\rangle:=(\partial_{\phi_j}\hat S^\dagger_{\bm{\xi}})|\bm{0}\rangle,
\end{equation}
where $|\tilde{1}_{s_j}\rangle\propto |0\rangle_L\otimes|b_1,\cdots,s_j,\cdots,b_N\rangle_A$ represents the state where only the $j$-th atom is excited from $|b\rangle_j$ to $|s\rangle_j$. Hence, we can deduce that
\begin{equation}
     \langle\bm{0}|\hat S'_{\bm{\xi}}{}^{n}\hat a^{m}|\tilde{1}_{s_j}\rangle=\mathrm{i}\frac{g^2_j}{\sum_k g_k^2}\delta_{n,1}\delta_{m,0}\delta_{s',s},
\end{equation}
with $\hat S_{\bm{\xi}},\hat S'_{\bm{\xi}}=\hat A_{\bm{\xi}}, \hat C_{\bm{\xi}}$ for $s,s'=a,c$. Therefore,
\begin{align}
\langle e(\tilde{m},\tilde{k},\tilde{n})|\tilde{1}_{s_j}\rangle
=&\,\delta_{\tilde{m},1}\delta_{\tilde{k},0}\delta_{\tilde{n},0}\langle\bm 0|\hat Q_{\bm{\xi},+}|\tilde{1}_{s_j}\rangle
+\delta_{\tilde{m},0}\delta_{\tilde{k},1}\delta_{\tilde{n},0}\langle\bm 0|\hat Q_{\bm{\xi},-}|\tilde{1}_{s_j}\rangle+\delta_{\tilde{m},0}\delta_{\tilde{k},0}\delta_{\tilde{n},1}\langle\bm 0|\hat D_{\bm{\xi}}|\tilde{1}_{s_j}\rangle\notag\\
=&\,\delta_{\tilde{m},1}\delta_{\tilde{k},0}\delta_{\tilde{n},0}\frac{1}{\sqrt{2}}(\langle\bm 0|\hat A_{\bm{ \xi}}|\tilde{1}_{s_j}\rangle
+\cos\theta_{\bm g}\langle\bm 0|\hat C_{\bm{\xi}}|\tilde{1}_{s_j}\rangle)+\delta_{\tilde{m},0}\delta_{\tilde{k},1}\delta_{\tilde{n},0}\frac{1}{\sqrt{2}}\big(\langle\bm 0|\hat A_{\bm{\xi}}|\tilde{1}_{s_j}\rangle\notag\\
&\,-\cos\theta_{\bm g}\langle\bm 0|\hat C_{\bm{\xi}}|\tilde{1}_{s_j}\rangle)-\delta_{\tilde{m},0}\delta_{\tilde{k},0}\delta_{\tilde{n},1}\sin\theta_{\bm g}\langle\bm 0|\hat C_{\bm{\xi}}|\tilde{1}_{s_j}\rangle\label{innerproduct_SM}\\
=&\,\mathrm{i}\frac{g_j^2}{\sum_k g_k^2}\{
\frac{1}{\sqrt{2}}\delta_{\tilde{m},1}\delta_{\tilde{k},0}\delta_{\tilde{n},0}(\delta_{a,s}+\cos\theta_{\bm g}\delta_{c,s})
+\frac{1}{\sqrt{2}}\delta_{\tilde{m},0}\delta_{\tilde{k},1}\delta_{\tilde{n},0}(\delta_{a,s}-\cos\theta_{\bm g}\delta_{c,s})\notag\\
&\,-\delta_{\tilde{m},0}\delta_{\tilde{k},0}\delta_{\tilde{n},1}\sin\theta_{\bm g}\delta_{c,s}\}\notag.
\end{align}
\subsection{Adiabatic condition and Berry's phase}\label{SubSec.~2_SM}
For Hamiltonian in Eq.~(\ref{Hamiltonian_initial_SM}), there exists a set of adiabatic variables $\bm\lambda=\{\phi_1,\cdots,\phi_N,\theta_{\bm{g}}\}$, with $\phi_j(t):=\Delta_j t$. The adiabatic condition required for a perfect adiabatic evolution of the quantum state $\sum_n C_n|D_{\bm{\xi},n}\rangle$ is expressed as
\begin{equation}
    \big|\frac{\langle e_{\bm{\xi}}(m',k',n')|\partial_t|D_{\bm{\xi},n}\rangle}{(m'-k')\varepsilon-0}\big|=\big|\frac{\langle e_{\bm{\xi}}(m',k',n')|\partial_{\bm{\lambda}}|e_{\bm{\xi}}(0,0,n)\rangle\cdot\dot{\bm{\lambda}}(t)}{(m'-k')\varepsilon-0}\big|\ll1.\label{adiabatic_condition_ini_SM}
\end{equation}

It follows from Eq.~(\ref{partial_theta_adiabatic_SM}) that
\begin{equation}
    \langle e_{\bm{\xi}}(m',k',n')|\partial_{\theta_{\bm{g}}}|e_{\bm{\xi}}(0,0,n)\rangle=-\sqrt{\frac{n}{2}}(\delta_{m',1}\delta_{k',0}-\delta_{m',0}\delta_{k',1})\delta_{n',n-1}.\label{adiabatic_partial_theta_inner_SM}
\end{equation}
Moreover, based on Eq.~(\ref{partial_phi_Dn_SM}) and Eq.~(\ref{innerproduct_SM}), we have
 \begin{align}
    \langle e_{\bm{\xi}}(m',k',n')|\partial_{\phi_j}|e_{\bm{\xi}}(0,0,n)\rangle=&\,\frac{1}{\sqrt{n!}}\langle e_{\bm{\xi}}(m',k',n')|\partial_{\phi_j}(\hat D^\dagger_{\bm{\xi}})^{n}|\bm{0}\rangle\notag\\
    =&\,\frac{n}{\sqrt{n!}}\langle e_{\bm{\xi}}(m',k',n')|(\hat D^\dagger_{\bm{\xi}})^{n-1}\partial_{\phi_j}(\hat D^\dagger_{\bm{\xi}})|\bm 0\rangle\notag\\
    =&\,-n[\frac{n'!}{n! (n'-n+1)!}]^{1/2}\sin\theta_{\bm{g}}\langle e_{\bm{\xi}}(m',k',n'-n+1)|\tilde{1}_{c_j}\rangle\Theta(n'-n+1)\label{adiabatic_partial_phi_inner_SM}\\
    =&\,-\mathrm{i}\frac{g_j^2}{\sum_k g_k^2}\sin\theta_{\bm g}\{\sqrt{\frac{n}{2}}\cos\theta_{\bm g}\delta_{n',n-1}(\delta_{m',1}\delta_{k',0}-\delta_{m',0}\delta_{k',1})\notag\\&\,-n\sin\theta_{\bm g}\delta_{m',0}\delta_{k',0}\delta_{n',n}\}\notag,
\end{align}
where $\Theta(\cdot)$ is the Heaviside function, defined by $\Theta(j)=1$ for $j\geq0$ and $\Theta(j)=0$ for $j<0$.

Therefore, by substituting Eqs.~(\ref{adiabatic_partial_theta_inner_SM}) and (\ref{adiabatic_partial_phi_inner_SM}) into Eq.~(\ref{adiabatic_condition_ini_SM}), the adiabatic condition can be written as
\begin{align}
    \frac{\sin\theta_{\bm{g}}}{(\sum_jg_j^2)^{1/2}}[\frac{n}{2}(|\dot\theta_{\bm{g}}|^2+|\langle\dot\phi\rangle_{\bm{\xi}}|^2\sin^2\theta_{\bm{g}}\cos^2\theta_{\bm{g}})]^{1/2}\ll1,
\end{align}
with
\begin{equation}
    \langle\dot\phi\rangle_{\bm{\xi}}:=\sum_j\dot\phi_j g_j^2/\sum_k g_k^2=\sum_j\Delta_j g_j^2/\sum_k g_k^2.
\end{equation}

Moreover, due to the consideration of $\Delta_j\sim\mathcal{N}(\Delta,\delta\Delta^2)$ and $g_j\sim\mathcal{N}(g,\delta g^2)$, we have $\sum_j g_j^2=N(\delta g^2+g^2)$. By expanding $\langle\dot\phi\rangle_{\bm{\xi}}$ around $g_j=g,\ \Delta_j=\Delta$, $\forall j$, we obtain $\langle\dot\phi\rangle_{\bm{\xi}}\simeq\sum_j\Delta_j/N\simeq\Delta$. Therefore, considering the stored state $|\phi_{\mathrm{in}}\rangle=\sum_n C_n|n\rangle_L$, the adiabatic condition can be further estimated as
\begin{equation}
    \tau_{\mathrm{d}},\ 1/|\Delta|\gg\sqrt{n_{\max}/[N(g^2+\delta g^2)]}\simeq\sqrt{n_{\max}/(Ng^2)},
\end{equation}
where we assume $\delta g\lesssim g$ and consider states with a finite photon number cutoff $n_{\max}$, such that $\sum_{n_{\max}+1}^\infty |C_n|^2\ll1$.

Furthermore, the Berry's phase is expressed as
\begin{equation}
    \gamma_{\bm{\xi},n}(\tau)=\mathrm{i}\int^\tau_0\mathrm{d}\bm\lambda\cdot\langle D_{\bm{\xi},n}|\partial_{\bm\lambda}|D_{\bm{\xi},n}\rangle,
\end{equation}
setting $m'=k'=0$, $n'=n$ in Eqs.~(\ref{adiabatic_partial_theta_inner_SM}) and (\ref{adiabatic_partial_phi_inner_SM}), and substituting into the above expression, we hence obtain
\begin{equation}
    \gamma_{\bm{\xi}}(\tau)=-n\langle\dot\phi\rangle_{\bm{\xi}}\int^\tau_0\mathrm{d}t\sin^2\theta_{\bm{g}}(t)=-n\sum_j\Delta_jg_j^2\int^\tau_0\mathrm{d}t\sin^2\theta_{\bm{g}}(t)/(\sum_k g_k^2).
\end{equation}

\subsection{High-order adiabatic process}
There exists a large zero-energy subspace $\mathcal{\bm{H}}_0:=\bigoplus_m\big\{ |e_{\bm{\xi}}(m,m,n)\rangle\big|\forall n\in\mathbb{Z}\big\}$ of the system in Eq.~(\ref{Hamiltonian_fin_SM}), however, the adiabatic condition does not forbid the transitions within the degenerate subspace $\mathcal{\bm{H}}_0$. Therefore, higher-order adiabatic approximations \cite{CPSun1990HighOrderAdiabatic} are required to determine whether transitions occur. Any state $|\varphi(t)\rangle=\sum_{mn}C_n^{m}|e_{\bm{\xi}}(m,m,n)\rangle$ in $\mathcal{\bm{H}}_0$ evolves according to
\begin{equation}
    \mathrm{i}\dot{C}^{[m]}_n(t)\simeq-\mathrm{i}\sum_{m',n'}\langle e_{\bm{\xi}}(m',m',n')|\partial_t|e_{\bm{\xi}}(m,m,n)\rangle C^{[m']}_{n'}(t),\label{high_order_adiabatic_equation_ini_SM}
\end{equation}
in cases of satisying adiabatic condition in Eq.~(\ref{adiabatic_condition_ini_SM}).

Based on Eqs.~(\ref{partial_phi_Dn_SM}) and (\ref{innerproduct_SM}), and similar to the derivation in Eq.~(\ref{partial_theta_adiabatic_SM}), we have 
\begin{align}
    \langle e_{\bm{\xi}}(m',m',n')|\partial_{\phi_j}|e_{\bm{\xi}}(m,m,n)\rangle=&\,\langle e_{\bm{\xi}}(m',m',n')|\{\frac{1}{\sqrt{m!}}\partial_{\phi_j}(\hat Q^\dagger_{\bm{\xi},+})^{m}|e_{\bm{\xi}}(0,m,n)\rangle+\frac{1}{\sqrt{m!}}\partial_{\phi_j}(\hat Q^\dagger_{\bm{\xi},-})^{m}|e_{\bm{\xi}}(m,0,n)\rangle\notag\\
    &\,+\frac{1}{\sqrt{n!}}\partial_{\phi_j}(\hat D^\dagger_{\bm{\xi}})^{n}|e_{\bm{\xi}}(m,m,0)\rangle\}\\
    =&\,m[\frac{m'!}{m!(m'-m+1)}]^{1/2}\{\langle e_{\bm{\xi}}(m'-m+1,m'-m,n'-n)|\frac{1}{\sqrt{2}}(|\tilde{1}_{a_j}\rangle+\cos\theta_{\bm{g}}|\tilde{1}_{c_j}\rangle)\notag\\
    &\,+\langle e_{\bm{\xi}}(m'-m,m'-m+1,n'-n)|\frac{1}{\sqrt{2}}(|\tilde{1}_{a_j}\rangle-\cos\theta_{\bm{g}}|\tilde{1}_{c_j}\rangle)\}\Theta(m'-m+1)\\&\,-n[\frac{n'!}{n!(n'-n+1)!}]^{1/2}\sin\theta_{\bm g}\langle e_{\bm{\xi}}(m'-m,m'-m,n'-n+1)|\tilde{1}_{c_j}\rangle\Theta(n'-n+1)\notag\\
    =&\,\mathrm{i}\frac{g_j^2}{\sum_k g_k^2}[m(1+\cos^2\theta_{\bm g})+n\sin^2\theta_{\bm g}]\delta_{m,m}\delta_{n,n'}.
\end{align}
Moreover, it follows from Eq.~(\ref{partial_theta_adiabatic_SM}) that $\langle e_{\bm{\xi}}(m',m',n')|\partial_{\theta_{\bm{g}}}|e_{\bm{\xi}}(m,m,n)\rangle=0$. Therefore, Eq.~(\ref{high_order_adiabatic_equation_ini_SM}) can be further written as
\begin{equation}
    \dot{C}^{[m]}_n(t)\simeq-\mathrm{i}[m(1+\cos^2\theta_{\bm g})+n\sin^2\theta_{\bm g}]\langle\dot\phi\rangle_{\bm{\xi}}{C}^{[m]}_n(t),
\end{equation}
since we choose $\big\{|D_{\bm{\xi},n}\rangle=|e_{\bm{\xi}}(0,0,n)\rangle\big|\forall n\big\}$ as storage space, i.e., $|\Phi(0)\rangle=\sum_n C_n|n\rangle_L\otimes|0\rangle_A=\sum_n C_n|D_{\bm{\xi},n}(0)\rangle$, we have $C_n^{[m]}(0)=C_n\delta_{m,0}$, therefore, it can be derived that $C^{[0]}_n(t)=C_n\exp(-\mathrm{i}n\langle\dot\phi\rangle_{\bm{\xi}}\int^t_0\mathrm{d}t'\sin^2\theta_{\bm{g}}(t'))$, which signifies Berry's phase factors. Moreover, the result $C^{[m\neq0]}_n(t)=0$ illustrates the fact that transitions between degenerate states do not occur.

\section{Review of quantum reliability and derivation of Eqs.~(\ref{reliability_1_letter}) and (\ref{reliability_2_letter})}\label{Appendix_Quantum_reliability}
In this Appendix, we first review the basic concept of quantum reliability \cite{reliability2023Cui} and compare it with standard fidelity measures in Appendix \ref{subAppendix_Quantum_reliability}. We then derive the reliability in Appendix \ref{subAppendix_Quantum_reliability_deviration} for the two procedures introduced in Sec.~\ref{Quantum reliability}, namely Eqs.~(\ref{reliability_1_letter}) and (\ref{reliability_2_letter}).

\subsection{brief review of quantum reliability \cite{reliability2023Cui} and comparison with fidelity}\label{subAppendix_Quantum_reliability}
In this sub-Appendix, we briefly review the concept of quantum reliability introduced in Ref.~\cite{reliability2023Cui} and compare it with standard fidelity measures, including state fidelity, process fidelity \cite{process_fidelity} and entanglement fidelity \cite{entanglement_fidelity1} Quantum reliability is a process-oriented metric based on the consistent quantum theory \cite{griffiths1984consistent_quantum_theory,griffiths2003consistent_)quantum_theory}. It characterizes the deviation between an actual evolution trajectory and a target trajectory. Therefore, it can be used as a trajectory-level extension of fidelity.

For a quantum system with Hilbert space $\mathcal{H}$, a physical property at a given time can be represented by a projection operator $\hat E$. More specifically, if the system state $|\psi\rangle$ satisfies $\hat E|\psi\rangle=|\psi\rangle$, then the state is said to have the property represented by $\hat E$. As an illustrated example, consider a three-level system $\{|0\rangle,|1\rangle,|2\rangle\}$. If the property of interest is that \textit{the system lies strictly in the subspace spanned by $|0\rangle$ and $|1\rangle$}, then this property is represented by $\hat E=|0\rangle\langle0|+|1\rangle\langle1|$. Whether the system has this property at a given time is called an \textit{event} at that time. In quantum reliability analysis, reliable events are usually defined according to whether the system lies in the subspace required to realize the desired function. Thus, $\hat E$ is chosen as the projector onto the reliable subspace, while $\hat E^\perp=\hat I-\hat E$ represents the unreliable subspace.

Furthermore, according to the consistent quantum theory \cite{griffiths1984consistent_quantum_theory,griffiths2003consistent_)quantum_theory}, if we consider a set of times $0<t_1<t_2<\cdots<t_f$, then a history, or equivalently a trajectory, is determined by a sequence of events at these times. Since either a reliable or an unreliable event may occur at each time, the system has $2^f$ possible trajectories. For example, the trajectory $|\psi_0\rangle\rightarrow {reliable}\rightarrow {unreliable}\rightarrow {reliable}\cdots \rightarrow {reliable}$ can be represented as $\mathcal{Y}=|\psi_0\rangle\langle\psi_0|\otimes \hat E_1\otimes \hat E^{\perp}_2\otimes \hat E_3\otimes\cdots\otimes \hat E_f$. For a closed system, the weight of this trajectory is defined as
\begin{equation}
    W[\mathcal{Y}]=\mathrm{Tr}[\hat E_f\hat U_f\cdots \hat E^{\perp}_2\hat U_2\hat E_1\hat U_1|\psi_0\rangle\langle\psi_0|\hat U_1^\dagger \hat E^\dagger_1\hat U_2^\dagger \hat E_2^{\perp\dagger}\cdots \hat U_f^\dagger \hat E_f^\dagger],
\end{equation}
where $\hat E_j$ and $\hat E_j^\perp=\hat I-\hat E_j$ denote the reliable and unreliable projectors at time $t_j$, respectively. Here, $\hat I$ is the identity operator, and $\hat U_j:=\hat U(t_j,t_{j-1})$ is the evolution operator from $t_{j-1}$ to $t_j$.

In reliability analysis, we are interested in whether the system remains reliable throughout the whole time interval. Therefore, comparing only the initial and final states is not sufficient. We also need to know whether a failure (unreliable) has occurred at an intermediate time. Once the system fails at a given time, the detailed state after this failure is usually no longer the object of interest. Thus, over the interval $[0,t_f]$, the family of histories relevant to reliability can be written as
\begin{align}
\mathcal{Y}_{\mathcal{F}_1}
=&\,|\psi_0\rangle\langle\psi_0|\otimes \hat E_1^{\perp}\otimes \hat I\otimes \cdots \otimes \hat I,\\
\mathcal{Y}_{\mathcal{F}_2}
=&\,|\psi_0\rangle\langle\psi_0|\otimes \hat E_1\otimes \hat E_2^{\perp}\otimes \hat I\otimes \cdots \otimes \hat I,\\
\cdots&\\
\mathcal{Y}_{\mathcal{F}_f}
=&\,|\psi_0\rangle\langle\psi_0|\otimes \hat E_1\otimes \hat E_2\otimes \cdots \otimes \hat E_f^{\perp},\\
\mathcal{Y}_{\mathcal{R}_f}
=&\,|\psi_0\rangle\langle\psi_0|\otimes \hat E_1\otimes \hat E_2\otimes \cdots \otimes \hat E_f .
\end{align}
Here, $\mathcal{Y}_{\mathcal{F}_j}$ denotes the history in which the system first fails at time $t_j$, while $\mathcal{Y}_{\mathcal{R}_f}$ denotes the survival history in which the system remains reliable at all considered times.

The reliability at time $t_f$ is defined as the weight of the survival history, $\mathcal{R}(t_f)=W(\mathcal{Y}_{\mathcal{R}_f})$, namely
\begin{equation}
    \mathcal{R}(t_f)=\mathrm{Tr}[
\prod_{i=1}^{f}(\hat E_i\hat U_i)
|\psi_0\rangle\langle\psi_0|
\prod_{i=f}^{1}(\hat U_i^\dagger \hat E_i^\dagger)
].
\end{equation}
This expression can be further generalized from a closed system to a more general setting. Suppose that the evolution in the $j$-th time interval is described by a quantum channel $\Lambda_j$. Then
\begin{equation}
    \mathcal{R}(t_f)=\mathrm{Tr}[
P_f\circ\Lambda_f\circ\cdots\circ P_1\circ\Lambda_1(\rho_0)
],
\end{equation}
where $\rho_0$ is the initial density matrix, and $P_j(\cdot)=\hat E_j(\cdot) \hat E_j^\dagger$ is the projection super-operator. These forms show that quantum reliability depends not only on the initial and final events (i.e., at $t=0$ and $t=t_f$), but also explicitly on the event structure at the intermediate times $\{t_1,\cdots,t_{f-1}\}$. Therefore, correlations between events at different times can naturally enter the reliability description.

It should be noted that a trajectory weight cannot always be interpreted directly as a classical probability. Such an interpretation is valid only when the corresponding family of histories satisfies the consistency condition. According to Ref.~\cite{reliability2023Cui}, one can introduce an appropriate measurement apparatus so that the relevant histories satisfy the consistency condition, while the corresponding trajectory weights remain invariant.

We now compare quantum reliability with fidelity. If only the initial and final events are considered, namely if the intermediate event structure is ignored, quantum reliability becomes $\mathcal{R}(t_f)=\mathrm{Tr}\left[\hat E_f\Lambda(\rho_0)\hat E_f^\dagger\right]$. Here, $\Lambda$ denotes the channel from the initial time to the final time. For a closed system, $\Lambda(\rho_0)=\hat U(t_f,0)\rho_0 \hat U^\dagger(t_f,0)$. If $\hat E_f=|\psi_{\rm tar}\rangle\langle\psi_{\rm tar}|$ is the projector onto a target pure state, then
\begin{equation}
    \mathcal{R}(t_f)=\langle\psi_{\rm tar}|\Lambda(\rho_0)|\psi_{\rm tar}\rangle,
\end{equation}
which is the usual state fidelity with respect to the target state. Thus, state fidelity essentially concerns only the initial and final events. It can be viewed as a special case of quantum reliability for a two-point history.

Entanglement fidelity \cite{entanglement_fidelity1} also has a two-point structure. For an input state $\rho$ and a channel $\Lambda$, the entanglement fidelity is defined as
\begin{equation}
    \mathcal{F}_e(\rho,\Lambda)=\langle\psi_\rho|
(I\otimes\Lambda)(|\psi_\rho\rangle\langle\psi_\rho|)
|\psi_\rho\rangle,
\end{equation}
where $|\psi_\rho\rangle$ is a purification of $\rho$. The quantity still compares the purified state before and after one application of the channel. It does not include the intermediate event structure inside the channel.

The same point applies to process fidelity. Although process fidelity is used to compare an actual channel process with an ideal one, it still has a two-point structure. The standard procedure is to represent the actual channel $\Lambda$ and the ideal channel $\Lambda_{\rm id}$ as quantum states \cite{process_fidelity}. For comparison, we assume here that the ideal channel is unitary, $\Lambda_{\rm id}(\cdot)=\hat U_{\rm Id}(\cdot)\hat U_{\rm Id}^\dagger$. The corresponding Choi states are defined as $\rho_\Lambda=\,(I\otimes\Lambda)(|\Phi\rangle\langle\Phi|)$ and $
|\Phi_{\Lambda_{\rm Id}}\rangle=\,(I\otimes \hat U_{\rm Id})|\Phi\rangle$
. Here, $|\Phi\rangle$ is usually chosen as the maximally entangled state $|\Phi\rangle=\sum_{n=0}^{d-1}|n\rangle\otimes|n\rangle/\sqrt{d}$. The process fidelity between the actual channel $\Lambda$ and the ideal channel $\Lambda_{\rm id}$ is then defined as
\begin{equation}
\mathcal{F}_{\rm pro}(\Lambda,\Lambda_{\rm id})
:=
\langle \Phi_{\Lambda_{\rm Id}}|\rho_\Lambda|\Phi_{\Lambda_{\rm Id}}\rangle
=
\langle \Phi_{\Lambda_{\rm Id}}|(I\otimes\Lambda)(|\Phi\rangle\langle\Phi|)|\Phi_{\Lambda_{\rm Id}}\rangle.
\end{equation}
Hence, process fidelity has a form similar to entanglement fidelity. It compares a selected state after the action of the full channel, and thereby indirectly compares the actual channel with the ideal one. It does not explicitly resolve the intermediate events inside the channel. In this sense, it is still a two-point history structure that focuses only on the initial and final states.

Therefore, quantum reliability can be used as a trajectory-level extension of fidelity. It not only concerns the initial and final events, but also retains the structural information associated with intermediate steps and their correlations, namely a multi-point history. Thus, in situations involving multiple processes, or multiple time intervals in which specific functions must be realized, it may provide a finer characterization of system degradation.

\subsection{Derivation of Quantum Reliability for the Procedures in Sec.~\ref{Quantum reliability}}\label{subAppendix_Quantum_reliability_deviration}

To facilitate the following discussion and derivation of the quantum reliability of these procedures, we first introduce several notations and preliminary results. 

It follows from Eq.~(\ref{Berry's_phase_expansion_letter}) that the output state approximately depends only on the collective detuning $\mathcal{D}:=\sum_j\Delta_j$ and the collective coupling strength $\mathcal{G}:=\sum_j g_j$. Therefore, for a given inhomogeneous configuration $\bm \xi$, we denote the corresponding single SIRO process as
\begin{equation}
    |\phi_{\mathrm{out}, \bm \xi}\rangle:=\hat U_{\Xi}(\tau)|\phi_{\mathrm{in}}\rangle,
\end{equation}
where the subscript $\Xi:=\{\mathcal{D},\mathcal{G}\}$. Moreover, based on the central-limit theorem, $\mathcal{D}\sim\mathcal{N}(N\Delta,N\delta\Delta^2)$ and $\mathcal{G}\sim\mathcal{N}(Ng,N\delta g^2)$.
\subsubsection{derivation of Eq.~(\ref{reliability_1_letter})}
In this subsection, we derive the quantum reliability of procedure (a), Eq.~(\ref{reliability_1_letter}), in detail. As shown in Fig.~\ref{fig2_paper} (a), the quantum memory performs SIRO processes on a sequence of $k$ states $\{|\phi_{\mathrm{in}}^{(1)}\rangle,\cdots,|\phi_{\mathrm{in}}^{(k)}\rangle\}$, with $|\phi^{(j)}_{\mathrm{in}}\rangle=\sum_n C_n^{(j)}|n\rangle_L$. 

Based on the framework of quantum reliability \cite{reliability2023Cui,reliability2025Cui,reliability2025Du,reliability2025Du2}, the reliable trajectory for a given configuration $\bm\xi$ can be written as
\begin{equation}
    |\phi_{\mathrm{in}}^{(1)}\rangle-\hat U_{\Xi}(\tau_1)\rightarrow reliable\dashrightarrow|\phi^{(2)}_{\mathrm{in}}\rangle-\hat U_{\Xi}(\tau_2)\rightarrow reliable\dashrightarrow\cdots,\label{trajectory_procedure_a}
\end{equation}
where $\tau_j$ represents the synchronization time for the $j$-th state, and the dashed lines indicate the intermediate process of retrieving out one state and storing the next one.

In this work, we focus only on the input and output times of each SIRO process, namely $\{t_1,t_1+\tau_1,\cdots,t_k-\tau_k,t_k\}$. According to the ideal function of a synchronizer, we define the \textit{reliable} event as the complete recovery of each input state $|\phi_{\mathrm{in}}^{(j)}\rangle$ after its SIRO process. Thus, the reliable event for the $j$-th SIRO process is represented by the projector $\hat {\bm E}^{(j)}:=|\phi_{\mathrm{in}}^{(j)}\rangle\langle \phi_{\mathrm{in}}^{(j)}|$. Since we focus on the reliability of SIRO processes indicated by the solid arrow, the above trajectory is equivalent to $|\Phi_{\mathrm{in}}\rangle-\bm{\hat U}_{\Xi}\rightarrow reliable$, with $|\Phi_{\mathrm{in}}\rangle:=\prod_j^k\otimes|\phi_{\mathrm{in}}^{(j)}\rangle$ and $\bm{\hat U}_{\Xi}=\prod_j^k\otimes{\hat U}_{\Xi}(\tau_j)$. The corresponding \textit{reliable} event is represented by $\bm \hat E=|\bm\Phi_{\mathrm{in}}\rangle\langle\bm\Phi_{\mathrm{in}}|$. For simplicity, we take $\tau_j=\tau$, $\forall j$. The quantum reliability of this procedure, namely the weight of the reliable trajectory, is then given by
\begin{align}
    \mathcal{R}_a(k)=&\,\langle\mathrm{Tr}[\hat{\bm{ E}}\bm{\hat U}_{\Xi}|\Phi_{\mathrm{in}}\rangle\langle\Phi_{\mathrm{in}}|\bm{\hat U}_{\Xi}^\dagger\hat{\bm{E}}^\dagger]\rangle_{\Xi}\\
    =&\,\sum_{\bm n,\bm n'}|C(\bm n)|^2|C(\bm n')|^2\exp[\mathrm{i}(\bm n-\bm n')\bm 1^{\mathrm{T}}\gamma_{\bm\xi_0,1}(\tau)]\exp[-(\bm n-\bm n')^{\mathrm{T}}\bm 1{\bm 1}^{\mathrm{T}}(\bm n-\bm n')\Gamma_{\bm\xi_0}(\tau)/(2N)],
\end{align}
where
\begin{equation}
    \Gamma_{\bm{\xi}_0}(\tau)=(\delta\Delta/\Delta)^2\gamma_{\bm\xi_0,1}^2(\tau)+(\delta g/g)^2\mu^2_{\bm{\xi}_0,1}(\tau),
\end{equation}
and $\bm{n}=(n_1,\cdots,n_k)^{\mathrm{T}}\in\mathbb{N}^k$, $\bm 1=(1,\cdots,1)^{\mathrm{T}}$, $C(\bm n)=\prod_j^kC^{(j)}_{n_j}$. Here, $\langle\cdot\rangle_{\Xi}$ represents averaging over random variable $\Xi$, which is equivalent to averaging over all random $\Delta_j$ and $g_j$.

\subsubsection{derivation of Eq.~(\ref{reliability_1_letter})}
As shown in Fig.~\ref{fig2_paper}(b), procedure (b) describes an arbitrary optical state $|\phi_{\mathrm{in}}\rangle$ undergoing $k$ SIRO processes in different quantum memories. As in the above discussion of procedure (a), we focus only on the input and output times of each SIRO process. The corresponding reliable trajectory is 
\begin{equation}
    |\phi_{\mathrm{in}}\rangle- \hat U_{\Xi_1}(\tau_1)\rightarrow reliable\cdots-\hat U_{\Xi_k}(\tau_k)\rightarrow reliable,
\end{equation}
where the subscript $\Xi_j$ corresponds to $j$-th quantum memory. Similarly, we define the \textit{reliable} event of this procedure as the complete recovery of the quantum state $|\phi_{\mathrm{in}}\rangle$ after it passes through the different memories. This \textit{reliable} event is represented by the projector $\hat E=|\phi_{\mathrm{in}}\rangle\langle\phi_{\mathrm{in}}|$. The quantum reliability of this procedure is therefore
\begin{align}
    \mathcal{R}_b(k)=&\,\langle\mathrm{Tr}[\prod^k_{j=1}\hat E\hat U_{\Xi_j} (\tau)|\phi_{\mathrm{in}}\rangle\langle\phi_{\mathrm{in}}|\prod^k_{j=1}\hat U^{\dagger}_{\Xi_j}(\tau)\hat E^\dagger]\rangle_{\Xi_1,\cdots,\Xi_k}\notag\\
    =&\,\sum_{\bm n,\bm n'}|C(\bm n)|^2|C(\bm n')|^2\cos[\gamma_{\bm{\xi}_0,1}(\tau)\bm 1^{\mathrm{T}}(\bm n-\bm n')]\exp[-(\bm{n}-\bm{n'})^{\mathrm{T}}(\bm{n}-\bm{n'})\Gamma_{\bm{\xi}_0}(\tau)/(2N)],\notag
\end{align}
where $\langle\cdot\rangle_{\Xi_1,\cdots,\Xi_k}$ denotes averaging over all $\Xi_1,\cdots,\Xi_k$, and $C(\bm n)$ is replaced with $\prod_j^k C_{n_j}$. Here the collective coupling and detuning associated with different quantum memories are treated as independent, since there is no direct interaction between each quantum memories.

\section{Origin and Definition of the Adiabatic-Pulse Factors $\kappa_\theta$ and $\zeta_\theta$.}\label{kappa_zeta}
In this Appendix, we briefly derive the adiabatic pulse factors $\kappa_\theta$ and $\zeta_\theta$ used in the paper.

We first consider $\kappa_\theta$. This factor comes from the phase $\gamma_{\bm{\xi}_0,1}(\tau)$ discussed in Eq.~(\ref{tomography_letter}) . To make the roles of the storage time $\tau_{\mathrm{s}}$ and the driving time $\tau_{\mathrm{d}}$ explicit, we separate out from $\gamma_{\bm{\xi}_0,1}(\tau)$. In our protocol, during the driving interval $t\in[0,\tau_{\mathrm{d}}/2]$, the driving field $\Omega(t)$ is adiabatically tuned from $\Omega\gg(\sum_j g_j^2)^{1/2}\simeq\sqrt{N}g$ to $\Omega\ll(\sum_j g_j^2)^{1/2}\simeq\sqrt{N}g$. Correspondingly, the angle $\theta_{\bm g_0}(t)$ changes from $\theta_{\bm g_0}\simeq0$ to $\theta_{\bm g_0}\simeq\pi/2$. During the storage interval $t\in[\tau_{\mathrm{d}}/2,\tau_{\mathrm{d}}/2+\tau_{\mathrm{s}}]$, one has $\theta_{\bm g_0}\simeq\pi/2$. The retrieve-out process takes place in the interval $t\in[\tau_{\mathrm{d}}/2+\tau_{\mathrm{s}},\tau]$, with $\tau=\tau_{\mathrm{d}}+\tau_{\mathrm{s}}$. As shown in Fig.~\ref{fig1_letter}, the retrieve-out pulse is taken to be symmetric with the store-in pulse, i.e., $\Omega(t)=\Omega(\tau-t)$, or equivalently $\theta_{\bm{g}_0}(t)=\theta_{\bm g_0}(\tau-t)$. It then follows that
\begin{align}
    \gamma_{\bm{\xi}_0,1}(\tau)=&\,-\Delta\int^{\tau}_0\mathrm{d}t\sin^2\theta_{\bm g_0}(t)=-\Delta[\int^{\tau_{\mathrm{d}}/2}_0\mathrm{d}t\sin^2\theta_{\bm g_0}(t)+\int^{\tau_{\mathrm{d}}/2+\tau_\mathrm{s}}_{\tau_{\mathrm{d}}/2}\mathrm{d}t\times1+\int^{\tau}_{\tau_{\mathrm{d}}/2+\tau_\mathrm{s}}\mathrm{d}t\sin^2\theta_{\bm g_0}(t)]\notag\\
    =&\,-\Delta[\int^{\tau_{\mathrm{d}}/2}_0\mathrm{d}t\sin^2\theta_{\bm g_0}(t)+\tau_\mathrm{s}-\int_{\tau_{\mathrm{d}}/2}^0\mathrm{d}t\sin^2\theta_{\bm g_0}(t)]:=-\Delta(\tau_{\mathrm{s}}+\kappa_\theta\tau_{\mathrm{d}}).
\end{align}
The dimensionless factor $\kappa_\theta$ is therefore defined as
\begin{equation}
    \kappa_\theta:=2\int^{1/2}_0\mathrm{d}(t/\tau_{\mathrm{d}})\sin^2\tilde{\theta}_{\bm g_0}(t/\tau_{\mathrm{d}}),
\end{equation}
where $\tilde{\theta}_{\bm g_0}(t/\tau_{\mathrm{d}}):=\theta_{\bm{g}_0}(t)$ is the function of the dimensionless variable $t/\tau_{\mathrm{d}}$. Therefore, $\kappa_\theta$ does not depend on the total driving time $\tau_{\mathrm{d}}$. It is determined solely by the pulse waveform.

We next consider $\zeta_\theta$. This factor comes from $\mu_{\bm{\xi}_0,1}(\tau)$ in Eq.~(\ref{mu}). As above, we separate the storage-time and driving-time contributions. We have
\begin{align}
    \mu_{\bm{\xi}_0,1}(\tau)=&\,-\frac{1}{2}\Delta\int^{\tau}_0\mathrm{d}t\sin^2[2\theta_{\bm g_0}(t)]=-\frac{1}{2}\Delta\{\int^{\tau_{\mathrm{d}}/2}_0\mathrm{d}t\sin^2[2\theta_{\bm g_0}(t)]+\int^{\tau_{\mathrm{d}}/2+\tau_\mathrm{s}}_{\tau_{\mathrm{d}}/2}\mathrm{d}t\times0+\int^{\tau}_{\tau_{\mathrm{d}}/2+\tau_\mathrm{s}}\mathrm{d}t\sin^2[2\theta_{\bm g_0}(t)]\}\notag\\
    =&\,-\Delta\int^{\tau_{\mathrm{d}}/2}_0\mathrm{d}t\sin^2[2\theta_{\bm g_0}(t)]:=-\Delta\zeta_\theta\tau_{\mathrm{d}},
\end{align}
with 
\begin{equation}
    \zeta_\theta:=\int^{1/2}_0\mathrm{d}(t/\tau_{\mathrm{d}})\sin^2[2\tilde{\theta}_{\bm g_0}(t/\tau_{\mathrm{d}})].
\end{equation}
Here the storage interval $t\in[\tau_{\mathrm{d}}/2,\tau_{\mathrm{d}}/2+\tau_{\mathrm{s}}]$ does not contribute because $\theta_{\bm g_0}\simeq\pi/2$ and hence $\sin^2(2\theta_{\bm g_0})\simeq0$. In addition, $\zeta_\theta$ is also a dimensionless factor determined only by the pulse waveform.

It should be noted that, for Eq.~(\ref{Gamma_letter}), in the long storage regime $\tau_{\mathrm{s}}\gg\tau_{\mathrm{d}}$, the leading contribution to $\Gamma_{\bm{\xi}_0}(\tau)$ reads
\begin{equation}
    \Gamma_{\bm{\xi}_0}(\tau)\simeq\delta\Delta^2\tau_{\mathrm{s}}^2+\Delta^2\delta g^2\zeta_\theta^2\tau_{\mathrm{d}}^2/g^2,\label{Gamma_approx}
\end{equation}
it gives the main contribution used in the capacity-time trade-off provided by Eq.~(\ref{trade_off_letter}).

\section{The Residual Detuning}\label{Section 3_SM}
Regarding current works that utilize tomography to measure the detuning of storage system \cite{Tomography2008squeezed,Tomography2009,Tomography2013}, which employ the relationship $\bar\varphi=\Delta^{(\mathrm{de})}\tau_{\mathrm{storage}}^{(\mathrm{de})}$. For a direct comparison, we adopt their definition of storage time in our work, namely the time interval between turning on and off the control field $\Omega$, hence $\tau_{\mathrm{storage}}=\tau_{\mathrm{s}}+\tau_{\mathrm{d}}/2$ in this paper (these two expressions are indistinguishable in $\tau_{\mathrm{s}}\gg\tau_{\mathrm{d}}$). Combining the relationship with the precise relationship in Eq.~(\ref{tomography_letter}) which considers the Berry's phase, one obtains the measured detuning $\Delta^{(\mathrm{de})}=\Delta(\tau_{\mathrm{s}}^{(\mathrm{de})}+\kappa_\theta\tau_{\mathrm{d}}^{(\mathrm{de})})/\tau_{\mathrm{storage}}^{(\mathrm{de})}$. Therefore, the omission leads to a residual detuning
\begin{equation}
\delta^{(\mathrm{de})}=\Delta^{\mathrm{(de)}}-\Delta=\alpha_\theta\tau^{(\mathrm{de})}_\mathrm{d}\Delta/(\tau^{(\mathrm{de})}_\mathrm{s}+\tau^{(\mathrm{de})}_\mathrm{d}/2),
\end{equation}
with $\alpha_\theta:=\kappa_\theta-1/2$.

\section{Mathematical Formulation of the Conditions discussed in Sec.~\ref{Section5_letter}}\label{condition_mathematic}

In this Appendix, we provide a mathematical formulation of the two conditions imposed on stored states discussed in Sec.~\ref{Section5_letter}. These conditions specify, respectively, how strongly the stored state should be localized in the Fock basis and how the normalized photon-number variable $Z_{n,\langle\Delta n^2\rangle}$ should approach a stable distribution.

For a stored state $|\phi_{\mathrm{in}}\rangle=\sum_n C_n |n\rangle_L $ with average photon number $\langle n\rangle:=\langle \phi_{\mathrm{in}}|\hat a^\dagger \hat a|\phi_{\mathrm{in}}\rangle$ and photon-number fluctuation $\langle\Delta n^2\rangle:=\langle \phi_{\mathrm{in}}|(\hat a^\dagger \hat a)^2|\phi_{\mathrm{in}}\rangle-\langle \phi_{\mathrm{in}}|\hat a^\dagger \hat a|\phi_{\mathrm{in}}\rangle^2$, the two conditions are stated as follows.

\emph{condition (1)}-- $|C_n|^2$ follows light-tailed distribution, where
\begin{equation}
    \sum_{|n-\langle n\rangle|\geq u}|C_n|^2\leq C\exp(-\chi u^\delta),\; \exists\ C,\chi,\delta>0,\ \forall u>0.
\end{equation}

In other words, the decay behavior of $|C_n|^2$ as it deviates from $\langle n\rangle$ is analogous to exponential type decay or even faster.

\emph{condition (2)}-- The distribution of state exhibit weak convergence in response to $\langle\Delta n\rangle$. The condition is described as follows:

We define the normalized random variable
\begin{equation}
    Z_{n,\langle\Delta n^2\rangle}=(n-\langle n\rangle)/\langle\Delta n^2\rangle^{1/2},\notag
\end{equation}
with mean value $\langle Z_{n,\langle\Delta n^2\rangle}\rangle=\sum_n |C_n|^2Z_{n,\langle\Delta n^2\rangle}=0$, and variance $\langle\Delta Z_{n,\langle\Delta n^2\rangle}\rangle=1$, its distribution is
\begin{equation}
    F_{\langle\Delta n^2{\rangle}}(y)=\sum_{y\geq Z_{n,\langle\Delta n^2\rangle}}|C_n|^2.
\end{equation}
Then $\exists\ F_Z(y)$,
\begin{equation}
    \lim_{\langle\Delta n^2\rangle\rightarrow\infty} F_{\langle\Delta n^2\rangle}(y)=F_Z(y).
\end{equation}

\section{Proof of fluctuation dominance}\label{Section 4_SM}
In the appendix, we demonstrate that, under the two conditions stated in Sec.~\ref{Section5_letter} and formulated more explicitly in Appendix~\ref{condition_mathematic}, and $\langle\Delta n^2\rangle\gg1$, the approximation in Eq.~(\ref{fidelity_large_fluctuation_letter}) becomes valid, and we will derive the explicit form of $c_m$. 

Expanding Eq.~(\ref{fidelity_letter}) with $\cos[(n-n')\gamma_{\bm{\xi}_0,1}(\tau)]=1$, we here obtain
\begin{equation}
    \mathcal{F}=\sum_m\frac{(-1)^m}{m!}(\frac{\Gamma_{\bm{\xi}_0}(\tau)}{2N})^m\sum_{n,n'}|C_n|^2|C_{n'}|^2(n-n')^{2m}.\label{fidelity_expansion_SM}
\end{equation}
Therefore, we now define
\begin{equation}
    c_{m,\langle\Delta n^2\rangle}:=\sum_{n,n'}|C_n|^2|C_{n'}|^2(Z_{n,\langle\Delta n^2\rangle}-Z_{n',\langle\Delta n^2\rangle})^{2m},\label{c_m,delta_supplementary}
\end{equation}
with
\begin{equation}
    Z_{n,\langle\Delta n^2\rangle}:=(n-\langle n\rangle)/\langle\Delta n^2\rangle^{1/2},
\end{equation}
hence we have the relation
\begin{equation}
    \sum_{n,n'}|C_n|^2|C_{n'}|^2(n-n')^{2m}=2c_{m,\langle\Delta n^2\rangle}\langle\Delta n^2\rangle^m.
\end{equation}
As follows, we prove that under conditions given in Appendix \ref{condition_mathematic}, there exists finite coefficient $c_m<\infty$, such that $c_{m,\langle\Delta n^2\rangle}$ converges to $c_m$ as $\langle\Delta n^2\rangle\rightarrow\infty$. In standard terms \cite{olver2010nist,billingsley2013convergence}, we need to demonstrate $c_{m,\langle\Delta n^2\rangle}$ converges in moments.

Therefore, we first present the definition of uniform integrability. Afterward, we demonstrate that the sequence of random variables $\{(Z_{n,\langle\Delta n^2\rangle})^k\}$, $\forall\,k\in\mathbb{Z}$ is uniformly integrability under \emph{condition (1)} given in Appendix \ref{condition_mathematic}. The definition is expressed as follows.

\emph{Definition of uniform integrability}: A sequence of random variables $\{Y_{\sigma}\}$ is uniformly integrable if
\begin{equation}
    \lim_{M\rightarrow\infty}\sup_{\sigma>0}\mathbb E[{|Y_\sigma}|\cdot I(|Y_\sigma|> M)]=0,
\end{equation}
where $I(|Y_\sigma|> M)$ is the indicator function that equals $1$ when $|Y_\sigma|> M$ and $0$ otherwise. In other words, absolute integrability requires that the distribution exhibit a well-behaved decay, as the variable deviates from its central value.

Based on \emph{condition (1)}, we now present \emph{Lemma 1} to facilitate our demonstration.

\emph{Lemma 1}: When a sequence of random variables $\{Y_\sigma\}$ satisfies $P(|Y_\sigma|\geq u)\leq a\exp(-b u^c),\forall u>0,\sigma>0,\exists a,b,c>0$, then $\{Y_\sigma\}$ is uniformly integrable.

\emph{Proof of Lemma 1}: Based on
\begin{align}
    \mathbb E[{|Y_\sigma}|\cdot I(|Y_\sigma|> M)]=&\,\int^\infty_0 \mathrm{d}u P(|Y_\sigma|\cdot I(|Y_\sigma|> M)\geq u)=\int^M_0\mathrm{d}u P(|Y_\sigma|>M)+\int^\infty_M\mathrm{d}u P(|Y_\sigma|\geq u)\notag\\=&\, MP(|Y_\sigma|>M)+\int^\infty_M\mathrm{d}u P(|Y_\sigma|\geq u)\\\leq&\,Ma\exp(-bM^c)+\int^\infty_M a\exp(-bu^c)\mathrm{d}u\\=&\,Ma\exp(-bM^c)+\frac{a}{c}b^{-1/c}\Gamma(\frac{1}{c},bM^c),
\end{align}
where $P(|Y_\sigma|\geq u)$ represents the cumulative distribution function, and $\Gamma(s,z)=\int_x^\infty\mathrm{d}u\exp(-u)u^{s-1}=z^{s-1}e^{-z}(1+(s-1)\vartheta(z)/z)$ is upper incomplete gamma function, with $\vartheta\rightarrow1$ as $z\rightarrow\infty$ \cite{olver2010nist}. Therefore,
\begin{equation}
    \lim_{M\rightarrow\infty}\sup_{\sigma>0}\mathbb[{|Y_\sigma}|\cdot I(|Y_\sigma|> M)]\leq\lim_{M\rightarrow\infty}Ma\exp(-bM^c)+\lim_{M\rightarrow\infty}\frac{a}{bc}M^{1-c}e^{-bM^c}(1+\frac{1-c}{bcM^c}\vartheta(bM^c))\rightarrow 0,
\end{equation}
$\{Y_\sigma\}$ is uniformly integrable.

Therefore, based on the \emph{condition (1)}, we have
\begin{equation}
    P(|(Z_{n,\langle\Delta n^2\rangle})^k|\geq u)=P(|n-\langle n\rangle|\geq u^{1/k}\langle\Delta n^2\rangle^{1/2})\leq C\exp(-\chi\langle\Delta n^2\rangle^{\delta/2} u^{\delta/k}),\exists\, C,\chi,\delta>0.
\end{equation}
By comparison with \emph{Lemma 1}, it is evident that $\{(Z_{n,\langle \Delta n^2\rangle)})^k\}$ is uniformly integrable. 

Subsequently, we present \emph{Theorem of Moment Convergence} (see Theorem 3.5 in Ref.~\cite{billingsley2013convergence}) as

\emph{Theorem of Moment Convergence}: If a sequence of random variables $\{Y_\sigma\}$, with the corresponding sequence of distribution function $\{F_\sigma(y)\}$, satisfies uniform integrability and weak convergence (i.e., $\exists\, F(y),\, \lim_{\sigma\rightarrow\infty}F_\sigma(y)=F(y))$, then $\exists\,\mathbb{E}(Y)<\infty,\,\lim_{\sigma\rightarrow\infty}\mathbb{E}(Y_\sigma)=\mathbb{E}(Y)$.

Therefore, by synthesizing \emph{condition (2)} and \emph{Theorem of Moment Convergence}, we can conclude that for $\{(Z_{n,\langle\Delta n^2\rangle})^k\},j\in\mathbb{Z}$, when $\langle\Delta n\rangle\rightarrow \infty$,
\begin{equation}
    \mathbb{E}[(Z_{n,\langle\Delta n^2\rangle})^k]=\sum_n|C_n|^2(Z_{n,\langle\Delta n^2\rangle})^k\Longrightarrow\mathbb{E}[(Z_n)^k]=\int^\infty_{-\infty}y^k\mathrm{d}F_Z(y),
\end{equation}
where $F_Z(y)$ is a convergent distribution function of the \emph{condition (2)} in Appendix \ref{condition_mathematic}. Hence, there exists finite $c_m<\infty$, 
\begin{align}
    c_{m,\langle\Delta n^2\rangle}=\sum_{k=0}^{2m}\left( \begin{array}{c}
    2m \\
    k 
    \end{array}\right)\mathbb{E}[(Z_{n,\langle\Delta n\rangle})^k]\mathbb{E}[(Z_{n,\langle\Delta n\rangle})^{2m-k}]\Longrightarrow
    c_m:=\sum_{k=0}^{2m}\left( \begin{array}{c}
    2m \\
    k 
    \end{array}\right)\mathbb{E}[(Z_n)^k]\mathbb{E}[(Z_n)^{2m-k}].
    \end{align}
Then $\sum_{n,n'}|C_n|^2|C_{n'}|^2(n-n')^{2m}=2c_{m,\langle\Delta n\rangle}\langle\Delta n^2\rangle^{m}\rightarrow 2c_m\langle\Delta n^2\rangle^{m}$ in case of $\langle\Delta n\rangle\rightarrow\infty$. Notably, $c_m$ solely depends on $F_Z(y)$ and index $m$. For the convergent distribution $F_Z(y)$, its moments of all orders are fixed, hence, $c_m$ is independent of $\langle\Delta n^2\rangle$. In other words, $c_m$ is the intrinsic property of stored states.

Consequently, when $\langle\Delta n^2\rangle\gg1$, Eq.~(\ref{fidelity_expansion_SM}) can be approximated as
\begin{equation}
    \mathcal{F}\simeq\sum_{m=0}\frac{c_m}{m!}[-\Gamma_{\bm{\xi}_0}(\tau)\langle\Delta n^2\rangle/N]^m,\label{fluctuation_domninance_SM}
\end{equation}
which indicates that $\mathcal{F}$ depends only on $\langle\Delta n^2\rangle$.

It is worth further noting that, based on the properties of Taylor series, Eq.~(\ref{fluctuation_domninance_SM}) may converge only within a specific range of $\langle\Delta n^2\rangle$ (i.e., within radius of convergence). However, since $\mathcal{F}$ is a real-analytic function, the identity theorem for real-analytic function (see Corollary 1.2.5 in Ref.~\cite{krantz2002realanalytic}) ensures that $\mathcal{F}$ remains dependent only on $\langle\Delta n^2\rangle$ over the whole range.

\section{Derivation and detailed form of fidelity for specific states}\label{Section 5_SM}
\subsection{Cat state}
In the subsection, we present the fidelity for a stored cat state as $|\phi_{\mathrm{in}}\rangle=\mathcal{N}_0(|\alpha\rangle+\exp(\eta+\mathrm{i}\theta)|-\alpha\rangle)$, $\forall\,\eta,\,\theta\in\mathbb{R}$, where
\begin{equation}
    \mathcal{N}_0=[1+\exp(2\eta)+2\exp(\eta-2|\alpha|^2)\cos\theta]^{-1/2},
\end{equation}
therefore, the explicit form of $C_n$ (i.e., $|\phi_{\mathrm{in}}\rangle=\sum_nC_n|n\rangle_L$) as
\begin{equation}
    C_n=\mathcal{N}_0\exp(-|\alpha|^2/2)[1+(-1)^n\exp(\eta+\mathrm{i}\theta)]\alpha^n/\sqrt{n!}.\label{cat_state_c_n_SM}
\end{equation}
In addition, when $\eta\rightarrow\pm\infty$, $|\phi_{\mathrm{in}}\rangle$ reduces to the coherent state $|\mp\alpha\rangle$.

We define
\begin{equation}
    \mathcal{K}(z)=\exp[-\Gamma_{\bm{\xi}_0}(\tau)z^2/(2N)]\cos[z\gamma_{\bm{\xi}_0,1}(\tau)],
\end{equation}
hence the fidelity can be further derived as
\begin{align}
    \mathcal{F}=&\,\int^\infty_{-\infty}\mathrm{d}z\mathrm{d}z'(\sum_n|C_n|^2\delta(z-n))(\sum_{n'}|C_{n'}|^2\delta(z'-n'))\mathcal{K}(z-z')\notag\\
    =&\,(\frac{1}{2\pi})^2\int^\infty_{-\infty}\mathrm{d}z\mathrm{d}z'\int^\infty_{-\infty}\mathrm{d}ke^{\mathrm{i}kz}(\sum_n|C_n|^2e^{-\mathrm{i}kn})\int^\infty_{-\infty}\mathrm{d}k'e^{\mathrm{i}k'z'}(\sum_{n'}|C_{n'}|^2e^{-\mathrm{i}k'n'})\mathcal{K}(z-z')\notag\\
    =&\,(\frac{1}{2\pi})^2\int^\infty_{-\infty}\mathrm{d}k\mathrm{d}k'(\sum_n|C_n|^2e^{-\mathrm{i}kn})(\sum_{n'}|C_{n'}|^2e^{-\mathrm{i}k'n'})\int^\infty_{-\infty}\mathrm{d}z\mathrm{d}z'e^{\mathrm{i}(kz+k'z')}\mathcal{K}(z-z')\label{derivation_cat_state_fidelity_SM}\\
    =&\,(\frac{1}{2\pi})^2\int^\infty_{-\infty}\mathrm{d}k\mathrm{d}k'(\sum_n|C_n|^2e^{-\mathrm{i}kn})(\sum_{n'}|C_{n'}|^2e^{-\mathrm{i}k'n'})\int^\infty_{-\infty}\mathrm{d}ye^{\mathrm{i}(k-k')y/2}\mathcal{K}(y)\int^\infty_{-\infty}\mathrm{d}Ye^{\mathrm{i}(k+k')Y}\notag\\
    =&\,(\frac{N}{8\pi\Gamma_{\bm{\xi}_0}(\tau)})^{1/2}\int^\infty_{-\infty}\mathrm{d}k|\sum_n|C_n|^2e^{\mathrm{i}kn}|^2\{\exp[-\frac{N(k+\gamma_{\bm{\xi}_0,1}(\tau))^2}{2\Gamma_{\bm{\xi}_0}(\tau)}]+\exp[-\frac{N(k-\gamma_{\bm{\xi}_0,1}(\tau) )^2}{2\Gamma_{\bm{\xi}_0}(\tau)}]\}\notag,
\end{align}
During the derivation, $z$ and $z'$ are transformed into the center-of-mass coordinate $Y:=(z+z')/2$ and the relative coordinate $y:=z-z'$. Therefore, regarding Eq.~(\ref{cat_state_c_n_SM}), we have
\begin{align}
    \sum_n|C_n|^2e^{\mathrm{i}kn}=&\,|\mathcal{N}_0|^2e^{-|\alpha|^2}\{|1+e^{\eta+\mathrm{i}\theta}|^2\sum_{m=0}\frac{(|\alpha|^{2}e^{\mathrm{i}k})^{2m}}{(2m)!}+|1-e^{\eta+\mathrm{i}\theta}|^2\sum_{m=0}\frac{(|\alpha|^{2}e^{\mathrm{i}k})^{2m+1}}{(2m+1)!}\}\notag\\
    =&\,|\mathcal{N}_0|^2e^{-|\alpha|^2}\{(1+2e^\eta\cos\theta+e^{2\eta})\cosh(|\alpha|^2e^{\mathrm{i}k})+(1-2e^\eta\cos\theta+e^{2\eta})\sinh(|\alpha|^2e^{\mathrm{i}k})\}\\
    =&\,\mathcal{E}_+\exp(|\alpha|^2e^{\mathrm{i}k})+\mathcal{E}_-\exp(-|\alpha|^2e^{\mathrm{i}k}),\notag
\end{align}
here, we define
\begin{align}
    \mathcal{E}_+=&\,\frac{1}{2}|\mathcal{N}|^2e^{-|\alpha|^2}\{(1+2e^\eta\cos\theta+e^{2\eta})+(1-2e^\eta\cos\theta+e^{2\eta})\}\notag\\
    =&\,2|\mathcal{N}|^2e^{-|\alpha|^2}e^\eta\cosh\eta,\\
    \mathcal{E}_-=&\,\frac{1}{2}|\mathcal{N}|^2e^{-|\alpha|^2}\{(1+2e^\eta\cos\theta+e^{2\eta})-(1-2e^\eta\cos\theta+e^{2\eta})\}\notag\\
    =&\,2|\mathcal{N}|^2e^{-|\alpha|^2}e^\eta\cos\theta.
\end{align}
Hence,
\begin{align}
    |\sum_n|C_n|^2e^{\mathrm{i}kn}|^2=&\,\mathcal{E}_+^2\exp({2|\alpha|^2\cos k})+\mathcal{E}_-^2\exp({-2|\alpha|^2\cos k})+2\mathcal{E}_+\mathcal{E}_-\cos(2|\alpha|^2\sin k)\notag\\
    =&\,\mathcal{E}_+^2\sum_{\nu=-\infty}^\infty I_\nu(2|\alpha|^2)e^{\mathrm{i}\nu k}+\mathcal{E}_-^2\sum^\infty_{\nu=-\infty}I_\nu(-2|\alpha|^2)e^{\mathrm{i}\nu k}+2\mathcal{E}_+\mathcal{E}_-\sum_{\nu=-\infty}^\infty J_\nu(2|\alpha|^2)\cos(\nu k),
\end{align}
where $J_\nu(z)$ ($I_\nu(z)$) refers to the (modified) Bessel function of the first kind. Then we substitute the above formula into Eq.~(\ref{derivation_cat_state_fidelity_SM}), hence
\begin{equation}
    \mathcal{F}=\sum_{\nu=-\infty}^{\infty}\cos[\nu\gamma_{\bm{\xi}_0}(\tau)]\exp[-\nu^2\Gamma_{\bm{\xi}_0}(\tau)/(2N)]\{\mathcal{E}_+^2I_\nu(2|\alpha|^2)+\mathcal{E}_-^2I_\nu (-2|\alpha|^2)+2\mathcal{E}_+\mathcal{E}_-J_\nu(2|\alpha|^2)\}.\label{fidelity_cat_state_appendix}
\end{equation}
We further examine the case of $\cos[\gamma_{\bm{\xi}_0}(\tau)(n-n')]=1$ and $\langle\Delta n^2\rangle\gg1$ (i.e., $|\alpha|^2\gg1$). Therefore, $\mathcal{N}\simeq(1+e^{2\eta})^{-1/2}$ due to $e^{\eta-2|\alpha|^2}\ll1$ when $\eta\leq0$ and $e^{\eta-2|\alpha|^2}\ll e^{2\eta}$ when $\eta>0$. Hence we have $\mathcal{E}_+\simeq\exp(-|\alpha|^2)$ and $\mathcal{E}_-\simeq\exp(-|\alpha|^2)\cos\theta/\cosh\eta$.
In addition, the modified Bessel function satisfies $I_\nu(-z)=(-1)^\nu I_\nu(z)$, here we have
\begin{align}
    \mathcal{F}=&\,\exp(-2|\alpha|^2)\{(1+\frac{\cos^2\theta}{\cosh^2\eta})\sum_{\nu=-\infty}^{\infty}\exp(-(2\nu)^2\Gamma_{\bm{\xi}_0}(\tau)/(2N))I_{2\nu}(2|\alpha|^2)\notag\\&\,+(1-\frac{\cos^2\theta}{\cosh^2\eta})\sum_{\nu=-\infty}^{\infty}\exp(-(2\nu+1)^2\Gamma_{\bm{\xi}_0}(\tau)/(2N))I_{2\nu+1}(2|\alpha|^2)\label{fidelity_cat_SM}\\&\,+2\frac{\cos\theta}{\cosh\eta}\sum_{\nu=-\infty}^\infty \exp(-\nu^2\Gamma_{\bm{\xi}_0}(\tau)/(2N))J_\nu(2|\alpha|^2)\}.\notag
\end{align}
In particular, when the stored state is a coherent state $|\alpha\rangle$ (i.e., $\eta\rightarrow-\infty$), the above expression reduces to
\begin{equation}
    \mathcal{F}=\exp(-2|\alpha|^2)\sum_{\nu=-\infty}^{\infty}\cos[\nu\gamma_{\bm{\xi}_0}(\tau)]\exp[-\nu^2\Gamma_{\bm{\xi}_0}(\tau)/(2N)]I_\nu(2|\alpha|^2).\label{fidelity_coherent_SM}
\end{equation}
For simplicity, we define $\mathcal{Z}=2|\alpha|^2$, hence $\exp(-\mathcal{Z})I_\nu(\mathcal{Z})\simeq\exp(-\nu^2/(2\mathcal{Z}))/\sqrt{2\pi\mathcal{Z}}$ in the case of $\mathcal{Z}\gg1$. Moreover, $J_\nu(\mathcal{Z})\simeq\sqrt{2/(\pi \mathcal{Z})}\cos(\mathcal{Z}-\nu\pi/2-\pi/4)\ll I_\nu(\mathcal{Z})$. Therefore, under $|\alpha|\gg1$, Eq.~(\ref{fidelity_cat_SM}) is approximated as
\begin{align}
    \mathcal{F}\simeq&\,\frac{1}{\sqrt{2\pi}}\{(1+\frac{\cos^2\theta}{\cosh^2\eta})\sum_{\nu=-\infty}^{\infty}\frac{1}{\sqrt{\mathcal{Z}}}\exp(-(2\nu)^2(\frac{\Gamma_{\bm{\xi}_0}(\tau)}{2N}+\frac{1}{2\mathcal{Z}}))\notag\\&\,+(1-\frac{\cos^2\theta}{\cosh^2\eta})\sum_{\nu=-\infty}^{\infty}\frac{1}{\sqrt{\mathcal{Z}}}\exp(-(2\nu+1)^2(\frac{\Gamma_{\bm{\xi}_0}(\tau)}{2N}+\frac{1}{2\mathcal{Z}}))\}\notag\\
    \simeq&\,\frac{1}{\sqrt{2\pi}}\{(1+\frac{\cos^2\theta}{\cosh^2\eta})\int^\infty_{-\infty}\mathrm{d}\tilde{\nu}\exp(-(2\sqrt{\mathcal{Z}}\tilde{\nu})^2(\frac{\Gamma_{\bm{\xi}_0}(\tau)}{2N}+\frac{1}{2\mathcal{Z}}))\\
    &\,+(1-\frac{\cos^2\theta}{\cosh^2\eta})\int^\infty_{-\infty}\mathrm{d}\tilde{\nu}\exp(-(2\sqrt{\mathcal{Z}}\tilde{\nu}+1)^2(\frac{\Gamma_{\bm{\xi}_0}(\tau)}{2N}+\frac{1}{2\mathcal{Z}}))\}\notag\\
    =&\,\{1+2|\alpha|^2\Gamma_{\bm{\xi}_0}(\tau)/(2N)\}^{-1/2}\simeq\{1+2\langle\Delta n^2\rangle\Gamma_{\bm{\xi}_0}(\tau)/(2N)\}^{-1/2}\notag,
\end{align}
where $\tilde{\nu}=\nu/\sqrt{\mathcal{Z}}$, thus $\sum1/\sqrt{\mathcal{Z}}\simeq\int\mathrm{d}\tilde{\nu}$. In addition, $\langle\Delta n^2\rangle\simeq|\alpha|^2$ where $\langle\Delta n^2\rangle\gg1$.

Moreover, it is noteworthy that when $\langle\Delta n^2\rangle\Gamma_{\bm\xi_0}(\tau)/(2N)\gg1$, we can obtain $\mathcal{F}\simeq[2\langle\Delta n^2\rangle\Gamma_{\bm\xi_0}(\tau)/(2N)]^{-1/2}$, which exhibit a power-law decay.

\subsection{Uniform superposition state}
In the subsection, we present the fidelity for the case where the stored state is a uniform superposition state
\begin{equation}
    |\phi_{\mathrm{in}}\rangle=\sum_{n=0}^{N_c-1}\frac{1}{\sqrt{N}_c}|n\rangle_L,
\end{equation}
where the photon number fluctuation $\langle\Delta n^2\rangle=(N_c^2-1)/12$. We here directly calculate the fidelity in the regime of $\cos[\gamma_{\bm{\xi}_0}(\tau)(n-n')]=1$ and $\langle\Delta n^2\rangle\gg1$ (i.e., $N_c\gg1$). We have
\begin{align}
    \mathcal{F}=\frac{1}{N_c^2}\sum_{n,n'}^{N_c-1}\exp[-\frac{1}{2N}\Gamma_{\bm{\xi}_0}(\tau)(n-n')^2]\simeq\int^{1}_0\mathrm{d}\tilde{n}\mathrm{d}\tilde{n}'\exp[-\frac{N_c^2}{2N}\Gamma_{\bm{\xi}_0}(\tau)(\tilde{n}-\tilde{n}')^2],
\end{align}
where we define $\tilde{n}=n/N_c$ with $N_c\gg1$, hence $\sum 1/N_c\simeq\int\mathrm{d}\tilde{n}$. Subsequently, we transform the coordinates $\tilde{n}$, $\tilde{n}$ into $\tilde{m}=(\tilde{n}+\tilde{n}')/2$ and $\tilde{m}'=(\tilde{n}-\tilde{n}')/2$, therefore,
\begin{align}
    \mathcal{F}\simeq&\,\int^{1/2}_{-1/2}\mathrm{d}\tilde{m}'\int^{1-|\tilde{m}'|}_{|\tilde{m}'|}\mathrm{d}\tilde{m}\exp[-2N_c^2\Gamma_{\bm{\xi}_0}(\tau){\tilde{m}'}{}^{2}/N]|\frac{\partial(\tilde{n},\,\tilde{n}')}{\partial(\tilde{m},\,\tilde{m}')}|\\
    =&\,4\int^{1/2}_0(1-2\tilde{m}')\exp[-2N_c^2\Gamma_{\bm{\xi}_0}(\tau){\tilde{m}'}{}^{2}/N]\\
    =&\,\sqrt{\pi}\mathrm{erf}(x)/x-\{\exp(-x^2)-1\}/x^2,
\end{align}

with
\begin{equation}
    x:=N_c^2\Gamma_{\bm{\xi}_0}(\tau)/(2N)\simeq 6\langle\Delta n^2\rangle\Gamma_{\bm{\xi}_0}(\tau)/(2N).
\end{equation}
It is further noted that in the regime of $x\gg1$, $\mathrm{erf}(x)\simeq1-\exp(-x^2)/(\sqrt{\pi}x)$, hence
\begin{equation}
    \mathcal{F}\simeq(\pi/6)^{1/2}(\langle\Delta n^2\rangle\Gamma_{\bm{\xi}_0}(\tau)/N])^{-1/2},
\end{equation}
which indicate it also exhibits a power-law tail decay.

\section{The proof of the lower bound of the fidelity}\label{Section 6_SM}
In the appendix, we present the demonstration of inequality in Eq.~(\ref{lowe_bound_letter}). We first define random variable $\kappa=(n-n')^2=0,1,\cdots$ with $n,n'\in\mathbb{Z}$, and define effective normalized probability $\widetilde{P}_\kappa=\sum_{(n-n')^2=\kappa}|C_n|^2|C_{n'}|^2$, hence the expectation is written as
\begin{equation}
    \mathbb{E}(\kappa)=\sum_{\kappa=0}\widetilde{P}_{\kappa}\kappa=\sum_{\kappa=0}(\sum_{(n-n')^2=\kappa}|C_n|^2|C_{n'}|^2)\kappa=\sum_{n,n'}|C_n|^2|C_{n'}|^2(n-n')^2=2\langle\Delta n^2\rangle.
\end{equation}
Furthermore, we define $g(\kappa)=\exp(-\Gamma_{\bm{\xi}_0}(\tau)\kappa/(2N))$, here one can observe that $g(\kappa)$ is a convex function with $\mathrm{d}^2g(\kappa)/(\mathrm{d}\kappa^2)>0$. Therefore, based on Jensen's Inequality, we have $\mathbb{E}(g(\kappa))\geq g(\mathbb{E}(\kappa))$, this implies
\begin{equation}
    \sum_{\kappa}\widetilde{P}_k\exp(-\Gamma_{\bm\xi_0}\kappa/(2N))\geq\exp(-\Gamma_{\bm\xi_0}\langle\Delta n^2\rangle/N),
\end{equation}
here
\begin{align}
    \sum_{\kappa}\widetilde{P}_k\exp(-\Gamma_{\bm\xi_0}\kappa/(2N))=&\,\sum_{\kappa}(\sum_{(n-n')^2=\kappa}|C_n|^2|C_{n'}|^2)\exp(-\Gamma_{\bm\xi_0}\kappa/(2N))\notag\\=&\,\sum_{n,n'}|C_n|^2|C_n'|^2\exp(-\Gamma_{\bm\xi_0}(n-n')^2/(2N)),
\end{align}
hence,
\begin{equation}
    \sum_{n,n'}|C_n|^2|C_n'|^2\exp(-\Gamma_{\bm\xi_0}(n-n')^2/(2N))\geq\exp(-\Gamma_{\bm\xi_0}\langle\Delta n^2\rangle/N).
\end{equation}

\section{Capacity of high fidelity quantum memory and Trade-off for capacity and time}\label{Section 7_SM}
In this Appendix, we present a detailed derivation of the explicit form of the effective information-storage capacity for the quantum memory in the presence of disordered coupling and disordered detuning. We also provide the capacity-time trade-off in Eq.~(\ref{trade_off_letter}) in the high-fidelity regime.

In the main text, we use the single-letter coherent information as an effective characterization of the maximum amount of information that the quantum memory can reliably store \cite{holevo2019QI,nielsen2010QI}, reads
\begin{equation}
    \mathcal{C}=\max_{\rho\in\mathcal{H}_d}[S(\Lambda(\rho))-S(\mathcal{I}\otimes\Lambda(|\psi_\rho\rangle\langle\psi_\rho|))]:=\max_{\rho\in\mathcal{H}_d}I_c(\rho),\label{LSD capacity_SM}
\end{equation}
where $\mathcal{H}_d$ is the fully supported Hilbert space, in which any state can be stored in the quantum memory to satisfy the requirement of universality, and its dimension is $d$ in Fock representation. Here, the $\Lambda(\cdot)$ represents the quantum channel corresponding to a single SIRO process, $|\psi_\rho\rangle$ is a purification of $\rho$, and $S(\cdot)$ is the von Neumann entropy.

It should be emphasized that, the information-storage capacity introduced in this paper is different from the \textit{quantum capacity} in quantum information theory. \textit{Quantum capacity} is a more general quantity built from the single-letter coherent information. Its definition further includes a regularization structure and extends to the setting of multiple channel uses. This allows joint encoding and joint decoding over the larger global space associated with repeated uses of the channel. By contrast, our subsequent discussion and trade-off analysis focus only on a single SIRO process in a memory. Our purpose is to introduce a effective and simple quantity to characterize how disorders affect the information-storage capability of the memory. For this purpose, the single-letter coherent information, as a basic building block of \textit{quantum capacity}, is sufficient for the discussion in this paper. It serves as an effective \textit{quantum capacity} proxy, namely, an effective measure of the information-storage capacity of the memory.

We next derive the explicit form of this effective information-storage capacity in the simultaneous presence of disorder. For random coupling strength and detuning $\bm\xi=\{\bm\Delta,\bm g\}$, one obtains
\begin{equation}
    \Lambda(\rho)=\sum_{n,n'}\rho_{n,n'}M_{n,n'}|n\rangle\langle n'|,\label{Lambda_SM}
\end{equation}
with
\begin{equation}
    M_{n,n'}=\exp\{-(n-n')^2\Gamma_{\bm\xi_0}(\tau)/(2N)\},
\end{equation}
where $\Gamma_{\bm\xi_0}$ coincides with the definition given in Eq.~(\ref{Gamma_letter}).

It can be proven that the density matrix achieving $\max_{\rho\in\mathcal{H}_d}I_c(\rho)$ has a diagonal form $\rho^*=\sum_n p_n|n\rangle\langle n|$, with $\sum_n p_n=1$, here the explicit form of $p_n$ remains to be determined. A detailed proof is provided in next subsection. The capacity $\mathcal{C}$ is then obtained by further maximizing $I_c(\rho^*)$.

Therefore, the first term of $I_c(\rho^*)$ is calculated as $S(\Lambda(\rho^*))=-\sum_n p_n \ln p_n$. For the second term, noting that the corresponding purification of $\rho^*$ is given by $|\psi_{\rho^*}\rangle=\sum_{n=0}^{d-1}\sqrt{p_n}|n\rangle\otimes|n\rangle$, one obtains
\begin{equation}
    \mathcal{I}\otimes\Lambda(|\psi_{\rho^*}\rangle\langle\psi_{\rho^*}|)=\sum_{n,n'=0}^{d-1}F_{n,n'}|n,n\rangle\langle n',n'|,
\end{equation}
where $F_{n,n'}=M_{n,n'}\sqrt{p_n p_{n'}}$, hence, 
\begin{equation}
    S(\mathcal{I}\otimes\Lambda(|\psi_{\rho^*}\rangle\langle\psi_{\rho^*}|))=\sum_{k=0}^{d-1}\lambda_k(F)\ln{\lambda_k(F)},
\end{equation}
where $\lambda_k(F)$ denotes the eigenvalues of matrix $F$.

In the following, we focus on the high fidelity region, i.e., $\epsilon:=\Gamma_{\bm\xi_0}/(2N)\ll1/\langle\Delta n^2\rangle\sim 1/d^2$, which follows from Eqs.~(\ref{fidelity_large_fluctuation_letter}) and (\ref{lowe_bound_letter}). The matrix $F$ is then expanded to first order $F\simeq F^{(0)}-\epsilon F^{(1)}$, the corresponding elements are $F^{(0)}_{n,n'}=\sqrt{p_n p_{n'}}$, $\epsilon F^{(1)}_{n,n'}=\epsilon (n-n')^2\sqrt{p_n p_{n'}}\leq\epsilon d^2\sqrt{p_n p_{n'}}\ll F^{(0)}_{n,n'}$. Therefore, using first-order perturbation theory, one then obtains
\begin{align}
    \lambda_0(F)\simeq&\, 1-2\epsilon \mathrm{var}(n)+O(\epsilon^2),\\
    \lambda_1(F)\simeq&\,2\epsilon \mathrm{var}(n)+O(\epsilon^2),\\
    \lambda_j(F)\simeq&\, O(\epsilon^2),\, j\geq2,
\end{align}
with
\begin{equation}
    \mathrm{var}(n):=\sum_np_n n^2-(\sum_n p_n n)^2.
\end{equation}

Hence, $I_c(\rho^*)$ in high fidelity regime is then obtained as
\begin{equation}
    I_c(\rho^*)\simeq-\sum_n p_n\ln p_n-2\epsilon \mathrm{var}(n)[\ln(2\epsilon \mathrm{var}(n))-1]+O(\epsilon^2).\label{capacity_dimension_relation_SM}
\end{equation}

Moreover, it can be proven that the dimension $d$ of Fock space $\mathcal{H}_d$ is related to the maximum photon number fluctuation of quantum states supported in the space via $(d-1)^2=4\langle\Delta n^2\rangle_{\max}$. It follows from Eq.~(\ref{fidelity_large_fluctuation_letter}) and (\ref{lowe_bound_letter}) that, in the high-reliable regime, the fidelity of the state with maximum photon number fluctuation after a SIRO process reaches the lower bound $\mathcal{F}_0$, with the relation given by
\begin{equation}
    1-\mathcal{F}_0=\langle\Delta n^2\rangle_{\max}\Gamma_{\bm{\xi}_0}/N.
\end{equation}
More specifically, when the photon number fluctuation of stored state satisfies $\langle\Delta n^2\rangle\leq\langle\Delta n^2\rangle_{\max}$, the fidelity $\mathcal{F}$ remains within the interval $[\mathcal{F}_0,1]$. Within the high reliable regime, one obtains $\Gamma_{\bm\xi_0}\langle\Delta n^2\rangle_{\max}/(2N)=\epsilon\langle\Delta n^2\rangle_{\max}=\epsilon(d-1)^2/4\ll1$. Moreover, the first term in Eq.~(\ref{capacity_dimension_relation_SM}) can attain its maximum value $\ln d$ for uniform distribution $p_n=1/d$. For the second term, since $\mathrm{vac}(n)\leq (d-1)^2/4$, it is much smaller than the first one, hence the storage capacity $\mathcal{C}$ can be well approximated in terms of the maximum allowable photon number fluctuation as
\begin{equation}
    \mathcal{C}\simeq\ln d=\ln\{(4\langle\Delta n^2\rangle_{\max})^{1/2}+1\}.
\end{equation}
When focusing on the regime $\langle\Delta n^2\rangle_{\max}\gg1$, one further obtains $\mathcal{C}\simeq1/2\ln(4\langle\Delta n^2\rangle_{\max})$. Therefore, combining Eq.~(\ref{Gamma_approx}), a general trade-off for storage capacity, storage and driving time is presented as
\begin{equation}
    1-\mathcal{F}_0\simeq(\delta\Delta^2\tau_{\mathrm{s}}^2+\Delta^2\delta g^2\zeta_\theta^2\tau_{\mathrm{d}}^2/g^2)(\exp\mathcal{C}-1)^2/(4N).
\end{equation}

\subsection{proof that the optimal density matrix is diagonal}
Before presenting the proof, we first state \emph{Lemma 2} as follows.

\emph{Lemma 2}: Let $\hat N=\hat a^{\dagger} \hat a$ be the number operator with eigen-states $|n\rangle$ spanning the Fock space $\mathcal{H}_d$, i.e., $\hat N|n\rangle=n|n\rangle$. The associated U(1) group is generated by group elements $\hat U(\phi)=\exp(\mathrm{i}\phi\hat N)$, $\forall \phi\in\mathbb{R}$. Then $I_c(\rho)$ in Eq.~(\ref{LSD capacity_SM}) is invariant under the action of U(1), i.e., $I_c(U(\phi)\rho U^\dagger(\phi))=I_c(\rho)$.

\emph{Proof of Lemma 2}: By defining $\rho_\phi=U(\phi)\rho U^{\dagger}(\phi)$ for $\forall\,\rho,\phi$, it follows from Eq.~(\ref{Lambda_SM}) that
\begin{equation}
    \Lambda(\rho_\phi)=\sum_{n,n'}\exp(\mathrm{i}\phi n)\rho_{n,n'}\exp(-\mathrm{i}\phi n')M_{n,n'}|n\rangle\langle n'|=U(\phi)\sum_{n,n'}\rho_{n,n'}M_{n,n'}|n\rangle\langle n'|U^{\dagger}(\phi)=U(\phi)\Lambda(\rho)U^{\dagger}(\phi).
\end{equation}
Since the von Neumann entropy is invariant under unitary conjugation, one obtains
\begin{equation}
    S(\Lambda(\rho_\phi))=S(U(\phi)\Lambda(\rho)U^{\dagger}(\phi))=S(\Lambda(\rho)).
\end{equation}

For the term $S(\mathcal{I}\otimes\Lambda(|\psi_\rho\rangle\langle\psi_\rho|))$, here $|\psi_\rho\rangle$ is the purification of $\rho$. A purification of $\rho_\phi$ can be given as
\begin{equation}
    |\psi_{\rho,\phi}\rangle=W\otimes U(\phi)|\psi_\rho\rangle,
\end{equation}
where $W$ is an arbitrary unitary operator. Therefore,
\begin{align}
    I\otimes\Lambda(|\psi_{\rho,\phi}\rangle\langle\psi_{\rho,\phi}|)=&\,I\otimes\Lambda[W\otimes U(\phi)|\psi_\rho\rangle\langle\psi_{\rho}|W^\dagger\otimes U^\dagger(\phi)]\notag\\
    =&\,(W\otimes I)(I\otimes\Lambda\{I\otimes U(\phi)|\psi_\rho\rangle\langle\psi_{\rho}|I\otimes U^\dagger(\phi)\})(W^\dagger\otimes I)\notag\\
    =&\,(W\otimes I)(I\otimes U(\phi)\Lambda(|\psi_\rho\rangle\langle\psi_{\rho}|) U^\dagger(\phi)\}(W^\dagger\otimes I)\\
    =&\,(W\otimes U(\phi))(I\otimes\Lambda(|\psi_\rho\rangle\langle\psi_\rho|))(W^\dagger\otimes U^\dagger(\phi)).\notag
\end{align}
Similarly,
\begin{equation}
    S(I\otimes\Lambda(|\psi_{\rho,\phi}\rangle\langle\psi_{\rho,\phi}|))=S[(W\otimes U(\phi))(I\otimes\Lambda(|\psi_\rho\rangle\langle\psi_\rho|))(W^\dagger\otimes U^\dagger(\phi))]=S(I\otimes\Lambda(|\psi_{\rho}\rangle\langle\psi_{\rho}|)).
\end{equation}
Therefore,
\begin{equation}
    I_c(U(\phi)\rho U^\dagger(\phi))=I_c(\rho).\label{Lemma2_SM}
\end{equation}

Furthermore, for channels of the type $\Lambda(\cdot)$ of Eq.~(\ref{Lambda_SM}), $I_c(\rho)$ is concave \cite{capacity_degerade2008}, as a discrete example, $I_c(\sum_j P_j \rho_j)\geq \sum_j P_jI_c(\rho_j)$, for $\forall \rho_j$ and  $P_j$. Therefore, for $\forall \rho\in\mathcal{H}_d$, the concavity implies that
\begin{equation}
    I_c(\int^{2\pi}_0\frac{\mathrm{d}\phi}{2\pi} \rho_{\phi})\geq \int^{2\pi}_0 \frac{\mathrm{d}\phi}{2\pi} I_c(\rho_\phi)=\int^{2\pi}_0\frac{\mathrm{d}\phi}{2\pi}I_c(U(\phi)\rho U^\dagger(\phi))=\int^{2\pi}_0\frac{\mathrm{d}\phi}{2\pi}I_c(\rho)=I_c(\rho),
\end{equation}
here, the \emph{Lemma 2} (i.e., Eq.~(\ref{Lemma2_SM}) is used and $\rho_\phi\in \mathcal{H}_d$, $\forall\phi$ holds as well. 

Hence, the above inequality must be saturated when maximizing $I_c(\rho)$ over $\mathcal{H}_d$, which indicates that the optimal density matrix has the form
\begin{equation}
    \rho^*=\int^{2\pi}_0\frac{\mathrm{d}\phi}{2\pi} \rho_{\phi}=\int^{2\pi}_0\frac{\mathrm{d}\phi}{2\pi}U(\phi)\rho U^\dagger(\phi)=\sum_{n,n'}\rho_{n,n'}\int^{2\pi}_0\frac{\mathrm{d}\phi}{2\pi}\exp(\mathrm{i}\phi(n-n'))|n\rangle\langle n'|=\sum_n p_n|n\rangle\langle n|,
\end{equation}
where $p_n:=\rho_{n,n}$ satisfies $\sum_n^{d-1} p_n=\mathrm{Tr}(\rho^*)=1$.

\end{widetext}

\bibliography{reference_letter}

\end{document}